# Photometry of LROC NAC resolved rock-rich regions on the Moon

Rachael M. Marshal[1], Ottaviano Rüsch[1], Christian Wöhler[2], Kay Wohlfarth[2], Sergey Velichko[3]

1. Institut für Planetologie, Westfälische Wilhelms Universität Münster, Münster, Germany
2. Image Analysis Group, Technical University of Dortmund, Dortmund, Germany
3. Institute of Astronomy of V.N. Karazin Kharkiv National University, Kharkiv, Ukraine

Email: rachael (.) marshal (at) uni-muenster (.) de

## 0. Abstract

The study and investigation of meter and sub-meter scale geological features, especially boulders and boulder fields, on the surface of airless bodies can provide insight into the evolution of the regolith and the contribution of various processes to its formation. Prior studies have examined the photometric properties of the lunar regolith surrounding young craters using image ratios. We extend this methodology to extracting surface properties, in particular the roughness characteristics, exclusive to boulder fields and the boulders that constitute them around impact craters. Our understanding of the response of boulders to space weathering, micrometeorite abrasion, thermal fatigue, and consequently their evolution into regolith can be improved by characterizing the surface roughness of the uppermost layer of boulders. In this study, rock-rich regions on the Moon are investigated using photometric roughness by employing a normalised logarithmic phase ratio difference metric to measure and compare the slope of the phase curve (reflectance versus phase angle) of a rock-rich field to a rock-free field. We compare the photometric roughness of rock-rich fields on simulated images with the photometric roughness of rock-rich fields on LROC NAC images (0.5m/pixel). The simulated terrains are constructed with a set of geologically informed rock properties, i.e., rock size, shape, and size frequency distribution. The artificial terrains are then converted to reflectance images using the Hapke AMSA (Hapke, 2002) reflectance model. Using this technique, we determine that rock-rich surfaces are not necessarily photometrically rougher than rock-free areas. Additionally, we find the roughness of resolved rock fields to indicate the presence of diverse sub-mm scale rock roughness (microtopography) and, possibly, variable rock single scattering albedo. These latter properties are likely controlled by rock petrology and material response to weathering and erosion. Spatial clustering of photometrically smooth and rough boulder fields in the downrange and uprange of two craters is observed, reflecting ejecta asymmetry and possibly indicating asymmetric modification of ejecta rock surfaces during the impact excavation process.

1. **Introduction**

Boulder fields are present on the lunar surface associated with impact craters: on the crater floor, slopes, and ejecta blankets, as well as perched on wrinkle ridges. Various studies have investigated the properties of boulder fields (in particular their spatial density and distribution) surrounding impact craters to derive information regarding their surface exposure age and regolith thickness (Bandfield et al., 2011; Ghent et al., 2014; Li et al., 2018; Watkins et al., 2019; Bickel et al., 2020; Rüsch et al., 2022). Key information can be derived on regolith formation and evolution by studying boulder fields since boulders constitute, as any ejecta material, the parent material of regolith. Due to their importance for regolith formation and evolution as well as for landing site selection, boulders constituting a boulder field, have been studied in terms of their morphometry (Noguchi et al., 2010; Rüsch & Wöhler, 2022), their spatial relationship to local scale relief (Krishna and Kumar, 2016; Watkins et al., 2019), their size-frequency distribution (SFD) (Hörz et al., 1975; Basilevsky et al., 2013; Watkins et al., 2019; Rüsch et al., 2022), and their abundance (e.g. Bandfield et al., 2011; Wei et al., 2020).

It is known that over time boulder fields experience erosion by impact shattering (e.g., Hörz et al., 2020; Rüsch et al., 2020). Individual rocks of a boulder field undergo weathering in the form of abrasion from micrometeorites (Hörz et al., 2020; Rüsch & Wöhler, 2022). Due to the effect of abrasion, the rocks develop a debris apron or a 'fillet' around them (Rüsch & Wöhler, 2022). Additionally, depending on the petrology, the boulder surface can undergo enhanced weathering due to diurnal temperature variation (Patzek and Rüsch, 2022). These changes in morphology and surface properties coupled with the decrease in boulder number with time and the associated change in boulder size frequency distribution are key to understanding the evolution of the boulder field and consequentially the regolith forming process (e.g., Basilevsky et al., 2013; Hörz et al., 2020; Rüsch et al., 2022).

By investigating the surface properties of boulder fields, in particular their roughness characteristics, it can be possible to gain insight into the evolution of the upper most surface of the rock that plays a key role in regolith formation.

This study aims to characterize boulder fields with photometric (optical) roughness. Photometry is usually studied to characterize rock-free regolith surfaces (e.g., Shepard & Campbell, 1998; Cord et al. 2007; Shepard & Helfenstein, 2007; Wu et al., 2009; Jin et al., 2015; Pilorget et al.,

2016; Labarre et al., 2017; Domingue et al., 2018 ; ) and occasionally to investigate rocks (e.g., Johnson et al., 2004; Tanabe et al. 2021; Hasselmann et al., 2021). Here rock photometry is investigated with the image ratio technique. This technique involves studying the surface properties of the area in consideration by using a ratio of two co-registered images acquired at varying viewing geometry (Buratti & Veverka, 1985; Shkuratov et al., 2011; Kaydash et al., 2012). The advantage of using phase ratio images is that the effects of varying albedo are largely cancelled out, and the resulting differences between terrains can be linked to surface roughness (Kaydash et al., 2012). Using the phase ratio method, photometric anomalies of anthropogenic and non-anthropogenic origin have been investigated. Previous studies (Kaydash et al., 2010; Kaydash et al., 2011; Kaydash & Shkuratov, 2012; Clegg et al., 2014; Clegg-Watkins et al., 2016; Velichko et al., 2022) show that anthropogenic activities have contributed to altering the surface roughness of the lunar regolith in the sub-mm scales in the following ways: (i) smoothening due to the fine-grained dust blown away by the gas jets, (ii) destruction of the "fairy castle" structure of the regolith which increases the brightness of the surface (Kreslavsky and Shkuratov, 2003) and iii) roughening as a result of regolith disturbance due to rovers, and astronaut boots. An outcome from the study of Kaydash & Shkuratov (2012) that is especially relevant to our study is the finding of an increase in surface roughness associated with the rays of a crater formed due to the impact of the Ranger 9 spacecraft. The reason for the increase surface roughness is attributed to rock fragments.

Photometric anomalies due to non-anthropogenic surface processes have also been studied using the phase ratio method. In their study, Kaydash et al. (2011), find an increase in photometric roughness on the rays of a small crater at the Apollo 14 site which is attributed to an abundance of rocks. Despite their advantages, phase ratio images are not immune to the effects of regional topography. For such reason, Velichko et al. (2020) employ a new technique to correct for these effects using the semi-empirical Akimov disk function. Subsequently, Velichko et al. (2020) use the topographically corrected data to characterise the heterogeneity of the ejecta blanket of a young crater on the inner flank of Hertzsprung S.

Based on the above-described studies, it is evident that multiple factors, including the presence of rocks, contribute to the photometric roughness of lunar images. In our study, we extend the phase ratio method in order to study and characterise the photometric roughness exclusive to boulders and boulder fields.

The approach of this study is to analyse the photometric behaviour, in particular the roughness, of resolved rock-rich surfaces via modelling and observation. In order to understand their behaviour, we investigate the Lunar Reconnaissance Orbiter Camera's Narrow Angle Camera (LROC NAC) data (Robinson et al., 2010) of resolved rock-rich surfaces using ratios of images acquired at varying viewing geometry. In resolved rock-rich surfaces the rock abundance and size distribution can be measured and allow us to better understand the different parameters that control the photometric behaviour. In particular, this study aims to disentangle the effect of rock morphology, abundance, size frequency distribution, and other parameters on the photometric response of a boulder field. For simplicity, in this study, we refer to the term rock independently of the size of the feature, from 1 mm to a few m wide. We refer to boulder fields as groups of resolved (>1 m in 0.5m resolution NAC images and >2m in 1m resolution NAC images) rocks identifiable in LROC NAC images.

## 2. Methods

### 2.1 Overview

We use artificial digital terrain models (DTMs) to simulate rock-rich surfaces and we estimate the reflectance of these artificial DTMs by employing the Hapke Anisotropic Multiple Scattering approximation (AMSA) model (Hapke, 2002; Grumpe et al., 2014). From the reflectance, we calculate a roughness metric (section 2.5) to quantify the resolved cm-m scale roughness of the artificial images. The same roughness metric is applied to NAC data for comparison purposes.

In the following sections, we present a detailed description of the procedure involved in creating the artificial DTMs, the details of the reflectance model and associated parameters, the roughness metric as well the procedure involved in selecting the resolved boulder fields from the NAC images. The model photometry is studied in two phases, first by analysing the results using a default set of parameter values, and second by varying each parameter independently to assess model sensitivity to input values.

### 2.2 Artificial Digital Terrain Model

The artificial digital terrain model (DTM) is a dimensionless horizontal flat area populated by rocks of variable abundance. The surface is populated randomly with the largest rocks being populated first, followed by smaller rocks. Rock abundance is defined throughout the study as the cumulative fractional area (CFA) of rocks. Naturally, rocks have a diversity of morphologies, aspect ratios, and size-frequency distributions (SFD) (as shown in figure 1). In this study, the goal is to investigate the roughness due to rock morphologies and abundances that are geologically plausible and occurring frequently, and not to model all possible morphological configurations.

Specifically, in our modelling approach, we use the rock aspect ratio from the study of Demidov & Basilevsky (2014) which determined the height to the maximum diameter of lunar rocks by analysing the panoramic images on board Lunokhod 1 and 2. In order to model rock morphologies shaped by micrometeorite abrasion, we use the rock topographic profiles (as well as debris apron profiles) that were determined by utilising the topographic diffusion equation in the study of Rüsch and Wöhler (2022). We use these previous studies in order for the model to be geologically informed.

There is a finite type of lunar rock petrology, mostly bimodal, and very few processes of rock formation and degradation acting on the Moon. Therefore, it can be expected that these morphologies do not vary drastically across the lunar surface. We note that this geologically informed topography represents a new approach to simulate surface roughness. Previous studies represented topographies using fractal statistics (e.g., Shepard & Campbell, 1998) or with sphere embedded in a surface to represent rocks (Shkuratov et al., 2005).

It is known that the SFD of a lunar rock population changes with time due to meteoroid bombardment (Rüsch et al., 2022; Hörz et al., 1975). The initial power-law distribution, usually the result of fragmentation, changes to an exponential distribution until only a few large survivor rocks are left. For the default modelling approach rocks follow an exponential SFD, typical for moderately mature rock populations (Rüsch et al, 2022; Hörz et al., 1975). In the second phase of the study we employ a second rock SFD to evaluate the effect of rock size on the roughness characteristic of a rock-rich region. The second rock SFD is described as two rocks of roughly the same size that are used to model a certain rock abundance, and represent a highly mature rock population. The exponential rock SFD is given by:

Exponential: $y = 1.5 * 0.15^{x*b}$          eq. (1)

Where in eq. 1, $y$ is the cumulative rock abundance and $x$ is the dimensionless rock size (diameter) and b is the exponent that is varied from 0.5 to 2.8 to obtain various cumulative rock abundances (CFAs). Variations of the exponent effectively slightly changes the shape of the SFD and allow us to model both small (<5%) and large (>60%) CFA. Diameters are binned with a constant increment. Rock shapes are modelled with an aspect ratio of c/a=0.57, and b=a, where a, b, and c are the first, second and third major axis and c is the height (Demidov & Basilevsky, 2014). This means that the height of rocks is defined by their (horizontal) width. As a consequence, the influence of rock height occurs only though a change in rock size, i.e., rock SFD.

For the rock topographic profiles, we use the rock shape descriptions presented in Rüsch & Wöhler (2022) that broadly correspond the Apollo-era rock shape classification into "angular", "subangular" and "round" (Muehlberger et al., 1972). We choose to discretize rock morphologies based on degradation level (figure 2). Non-degraded rocks are angular rocks modelled with top faces tilted at 0, 10°, 20°, and 30° angles. For each rock, the orientation of the tilted faces is across all azimuth angles. Despite this rare configuration, rocks usually present a single tilted top face, this approximation is sufficient since we are interested in the mean reflectance of the synthetic surface, and so the mean illumination and viewing geometry of all rocks present on the artificial surface. These shapes represent rocks that are not acted on by micrometeoroid abrasion and/or are composed of high-strength material (Rüsch & Wöhler, 2022). The angularity originates during the disruption of the parent material (bedrock or rock). Mildly and highly abraded rocks have increasingly subdued transitions between the top and side faces (figure 2) due to the process of micrometeoroid impacts rounding off edges as described in Rüsch & Wöhler (2022). The latter two rock types have debris aprons at their base of different heights. These last two rock shapes are indicative of the level of erosion owing to either their material strength and/or the surface exposure time.

The artificial DTM is composed of 500*500 cells. Each cell contains an elevation value and information on whether it constitutes a rock section, an underlaying regolith, or a debris apron.

## 2.3  Reflectance Model

The artificial surface is composed of regolith and rock areas. We assume that the debris apron has the same properties as the nearby regolith. Following the DTM creation, the local incidence and emission angles are calculated for each cell by ray tracing and these local angles are then input into the reflectance model.

To calculate regolith reflectance, we use the Anisotropic Multiple Scattering Approximation formulation of the Hapke reflectance model (Hapke 2002). The equation of the Hapke AMSA (Hapke, 2002) adaption is:

$$R_{AMSA}(\mu_\circ, \mu, \alpha) = \frac{\omega}{4\pi}\frac{\mu_\circ}{\mu_\circ+\mu}[p(\alpha)B_{SH}(\alpha) + M(\mu_\circ,\mu)]B_{CB}(\alpha) \qquad \text{eq. (2)}$$

where $\omega$ is the single scattering albedo (SSA), $\mu_\circ = cos(i)$ and $\mu = cos(e)$ where $i$ and $e$ are incidence and emission angles. The SSA is given by the ratio of the scattering coefficient of the medium in consideration to the extinction coefficient (Hapke, 1981).

Parameters $B_{SH}$ and $B_{CB}$ are the shadow-hiding and coherent backscatter functions. Together these functions describe the surge in brightness that occurs when imaging particulate media including regolith surfaces at very low phase angles (Hapke, 2012). In our study, we solely employ LROC NAC observations at phase angles >40°. At these large phase angles, the opposition effect is negligible and hence we ignore the correction terms for the opposition surge in equation (2) as done by Lin et al. (2020). Moreover, at phase angles greater than 40°, macroscopic roughness (the regolith surface topography at <mm–cm scales) are significant (e.g., Shkuratov et al., 2005; Shkuratov et al., 2012; Belgacem et al., 2021).

The single particle scattering function $p(\alpha)$ otherwise known as the single-particle phase function describes the nature of scattering of light by the individual particles in the medium (Hapke, 1993). The particle phase function $p(\alpha)$ is given by the double Henyey Greenstein (HG) function as follows:

$$p(\alpha) = \frac{1+c}{2}\frac{1-b^2}{(1+2bcos\alpha+b^2)^{\frac{3}{2}}} + \frac{1-c}{2}\frac{1-b^2}{(1-2bcos\alpha+b^2)^{\frac{3}{2}}} \qquad \text{eq. (3)}$$

where $\alpha$ is the phase angle, $b$ is the angular width of the scattering lobe and $c$ is the amplitude of the scattering lobe (Hapke, 1995). The relationship between the $b$ and $c$ values, through various experiments, is given by Hapke (2012) as:

$$c = 3.29exp(-17.4b^2) - 0.908 \qquad \text{eq. (4)}$$

The above relationship is often referred to as the hockey stick relation. From the experimental study by McGuire & Hapke (1995), particles with a large number of internal scatterers tend to be more backward scattering while transparent particles exhibit more forward scattering behaviour (Shepard & Helfenstein, 2007).

The term denoted by M refers to the multiple scattering function (Hapke, 2002) and is given by

$$M(\mu_o,\mu) = P(\mu_o)[H(\mu) - 1] + P(\mu)[H(\mu_o) - 1] + P[H(\mu) - 1][H(\mu_o) - 1] \qquad \text{eq. (5)}$$

Where $P(\mu_o)$ is the average radiance scattered into the lower hemisphere by a particle at an angle *i* with respect to the normal, $P(\mu)$ is the average radiance scattered in the upper hemisphere at angle *e* with the vertical and *P* is denoted as the radiance scattered by a particle into the lower hemisphere when illuminated uniformly from the lower hemisphere (Hapke, 2002).

In the default modelling approach, we do not include additional dependency or corrections to the eq. (2) such as for macroscopic roughness (Hapke, 2012) or porosity (Hapke, 2008). This is because input parameters for these additional functions (filling factor, mean gaussian slope) are poorly constrained and partly coupled with other parameters (Sato et al., 2014; Schmidt & Fernando, 2015; Helfenstein & Shepard, 1999). Moreover, we are first interested in quantifying the roughness due to shadow-causing facets in the scale of the modelled DTMs rather than at sub-mm/particle scale. In a second phase of the modelling effort, in order to investigate the effect of shadow-causing facets well below the DTM scale (sub cm and mm scale), we do calculate the relative change in modelled photometric roughness when the correction for macroscopic roughness, i.e., $\bar{\theta}$, is included. The correction for macroscopic roughness is given by the shadowing function $S(\bar{\theta},\mu_o,\mu)$ which is described in Hapke (1984). The shadowing function S is multiplied linearly to equation 4. The macroscopic roughness $\bar{\theta}$ is defined as the mean slope angle and is an integration of all sub resolution shadow causing facets, where the smallest (sub mm scale) and roughest facets dominate (Hapke 1984; Helfenstein et al., 1988; Shepard & Helfenstein, 2007).

In general, solid rock surfaces display a diffusive behaviour similar to a Lambertian surface (Hapke & van Horn, 1963). Optically thin grains within slabs exhibit forward scattering behaviour (Hapke, 1993), however, when embedded in a matrix (i.e., when considered within the rock texture), silicate rocks exhibit rather backscattering behaviour (Sato et al., 2014). To accurately model the combined effect of dust and underlying rock, a two layers model presented in Hapke (1993) can be used (e.g. Johnson et al., 2004; Sun et al., 2022). In this formalism, an optically thin layer is superimposed on an optically thick substrate and the two layers have different single scattering albedo and scattering properties. For the purposes of our modelling, two endmember reflectance can be considered: a complete dust coverage and a dust-free case. The photometric behaviour of complete dust coverage would be identical to that of the regolith. In the dust-free case, we utilize the same formalism as eq. (2) with isotropic phase function for rock surfaces, because if multiple scattering is considered negligible, the reflectance function of particulate material and solid slab is identical. In this simplified case, the reflectance is solely a function of the single scattering albedo, the Lommel-Seeliger factor, and the phase function.

## 2.4 Reflectance Model parameters

In order to calculate reflectance values for each cell of the artificial surface, the input parameters to calculate reflectance are single scattering albedo $\omega$, and the parameter of the phase function, $b$ (Table 1). As these parameters are linked to the physical properties of the material, we chose to disentangle between $\omega$, $b$, and $c$ for cells belonging to rocks and to regolith/debris apron.

The $\omega$ parameter is dependent on the optical constants, i.e., the intrinsic properties of the material (composition) and grain size (e.g., Hapke 1981; Sun et al., 2021; Zhou et al., 2021). The rationale for a higher $\omega$ for rocks is based on laboratory reflectance measurements and previous modelling of lunar rock albedo. Adams & McCord (1971) as well as Clark et al. (2002) found that lunar rocks can have reflectances up to a few times the reflectance of soils. Accordingly, previous reflectance modelling of lunar rocks has set the albedo to about 1.5 times the albedo of the regolith (Bandfield et al., 2011; Rüsch &Wöhler, 2022). Recently, De Angelis et al. (2017) retrieved $\omega$ values for a basaltic slab measured in the laboratory. We therefore adopted this value as it is the closest representative for our case. Additionally, we see from Clark et al. (2002) that the Apollo 16 fragmental breccia rocks have a higher reflectance than the soil (almost ~3 times than the soil) at NAC wavelength. Our choice of higher albedo for rock than regolith results in rock reflectance ~2.5 higher than the regolith. With this choice, the obtained reflectance values are within the values of Clark et al. (2002).

For the regolith single scattering albedo of both study regions, i.e., North Ray and the unnamed crater at Hertzsprung S, we use the $\omega$ value calculated by Watkins et al. (2021). This value is representative of highland material and therefore a reasonable choice for both study regions. As mentioned earlier, since $\omega$ is related to the composition of the material and from previous studies (Helfenstein and Veverka, 1987; Clegg Watkins et al., 2014c; Jolliff et al., 2021), $\omega$ shows an increase with decreasing FeO content (mafic mineralogy). The mafic content of the ejecta blanket of North Ray as well as that of the unnamed crater at Hertzsprung S are comparable in terms of FeO and $TiO_2$ weight percentage (Hawke et al., 1991; Spudis et al., 1996; Fortezzo et al., 2020) thereby justifying our usage of identical $\omega$ the two study regions.

However, we remark that in general the association of $\omega$ to the intrinsic property of an individual particle is not confirmed since $\omega$ might differ depending on the packing state, as found by Shepard & Helfenstein (2007). In fact, $\omega$ is more accurately termed as volume single scattering albedo (Hapke, 1999). We highlight that the choice of these values is of secondary importance because in our trial to extract reflectance from our artificial DTMs, we do not invert the model and extract meaning from them - rather we use previously estimated values to calculate reflectance.

For the *b* and *c* parameters of the HG phase function, we choose values representative for highland mature soil estimate from the study of Sato et al. 2014, for both our study regions. This simple choice is justified by the lack of a comprehensive study on slab scattering properties.

Lastly, we assume that the reflectance of the shadowed regions is negligible. We do not consider the effect of multiple scattering due to topography (at the scale of the DTM) of the modelled surface since the albedo of the lunar surface is generally low and the multiple scattering contribution is minor (Shkuratov et al., 2005). An example of the created artificial images for various rock morphologies at ~25% rock abundance (exponential distribution) is shown in Figure 3. A brief description of all the utilised parameters in producing these artificial images is given in Table 1.

## 2.5  Normalised Log Phase Ratio Difference

Phase ratio images provide an indication of the photometric behaviour of a given surface (e.g., Shkuratov et al., 2012). In order to isolate the photometric properties of rocks we remove other effects (e.g., regolith properties in between rocks) by normalizing the phase ratio image with the phase ratio image of rock-free surfaces. To do so we employ the Normalised Log Phase Ratio Difference (NLPRD) calculated using the following equation:

$$NLPRD = \frac{log10\left(\frac{F_{P2}}{F_{P1}}\right) - log10\left(\frac{B_{P2}}{B_{P1}}\right)}{(P2 - P1)} \quad \text{eq. (6)}$$

Where F and B denote the mean reflectance of the reference rock-free area and rock-rich area (boulder field), respectively. In the case of the artificial images the reflectance (calculated using eq. 2) is the spatially averaged model reflectance. In the case of the NAC data, we convert the I/F to equigonal albedo or $A_{eq}$ (as described in section 2.6, and in Velichko et al. 2020). The mean $A_{eq}$ is calculated by averaging each NAC pixel within each area. The selection of the NAC areas is presented further below. $P_2$ and $P_1$ refer phase angles such that $P_2 > P_1$. Effectively, the $\frac{F_{P2}}{F_{P1}}$ and $\frac{B_{P2}}{B_{P1}}$ are phase ratio images. This parameter relies on the same principle employed in the study of Shkuratov et al. (2012), i.e., the phase ratio values are normalised to the phase ratio value averaged over the entire study area and at a given reflectance. The NLPRD is in fact proportional to the roughness parameter δ defined in Shkuratov et al. (2012). Positive values of NLPRD are indicative rock-rich surfaces being rougher than rock free surfaces, while negative values indicate the inverse.

## 2.6 Selection of LROC NAC images, rock-rich and rock free areas

In this study, we focus on two regions: near an unnamed young crater in the inner flank of Hertzsprung S (0.778°N, 132.95°W) and near North Ray crater (8.82°S,15.48°E). The first region is selected because of the availability of image pairs with suitable phase angles (Table 2) that have been topographically corrected in the study of Velichko et al. (2020). The unnamed crater is 700 m in diameter and exhibits many large rocks on its rim that are resolved in NAC images. The image pair have very similar solar latitude and longitude (approximately the same local time), very similar incidence angle, different emission angles, and different phase angles. With this geometry, which we, as other authors (e.g., Kaydash et al., 2012) refer to as "emission ratio", or "configuration 1", the possible influence of the incidence angle on the phase ratio values is avoided (Kaydash et al., 2012). The second region is selected because of the higher abundance of rocks resolved in NAC images and the availability of in situ data for ground truth acquired during the Apollo 16 mission. This region has abundant image pairs with more standard NAC image acquisition geometries, i.e., very similar solar latitude and longitude (approximately the same local time), emission close to nadir in both images, and different incidence (hence different phase angles), a configuration we refer to as "incidence ratio" and "configuration 2" (See Table 2). We note that phase image pairs with different solar latitude and longitude exist for this study area but are unsuitable due to the irregular regional scale topography.

Comparison between the results obtained with the two image pairs, emission and incidence image ratios, allow us to assess the influence of different geometric combinations within the sampled phase angle range (~40-70°) on the photometric behaviour of rock-rich surfaces. LROC NAC images are downloaded in the EDR (Experimental Data Record) format from the Lunar Orbital Data Explorer (https://ode.rsl.wustl.edu/moon/index.aspx) and consequently pre-processed (radiometric correction and echo removal), spatially referenced and resampled to the digital elevation model resolution of 1m/pixel using the ISIS3 software (Sides et al., 2017). The pre-processed NAC images were further corrected to reduce the effects of regional topography as elaborated in Shkuratov et al. (2011) in order to make them suitable for photometric studies.

As discussed by Velichko et al. (2020), it is necessary to correct the effects of topography in order to avoid variations in the albedo that may be unrelated to surface texture. To correct for regional slopes, a digital elevation model is needed. For the North Ray study region, we use a digital elevation model derived from LRO NAC images using the technique of Grumpe et al. (2014) with a resolution of 1m/pixel. In order to extract the $A_{eq}$ we use the equation given by Shkuratov et al. (2011) expressed as

$$A_{eq}(\alpha) = A(\alpha,\beta,\gamma)/D(\alpha,\beta,\gamma) \qquad \text{eq. (7)}$$

Where $\alpha$ is the phase angle and $\beta$ and $\gamma$ are the photometric latitude and longitude.

Similar to the workflow of Velichko et al. (2020) we also employ the semi-empirical Akimov disc function (Akimov, 1988a, 1988b; Akimov et al., 1999, 2000; Shkuratov et al., 2011; Velikodsky et al., 2011)

$$D(\alpha,\beta,\gamma) = \cos(\alpha/2) (\cos \beta)^{\nu\alpha/(\pi-\alpha)} \cos[(\gamma\ \alpha/2)\pi/(\pi-\alpha)]/\cos \gamma \qquad \text{eq. (8)}$$

In the above equation, $\nu$ refers to the sub resolution roughness and is given a value of 0.52 representative for highlands (Akimov et al., 1999, 2000).

For the study area Hertzsprung S, we use the data that was topographically corrected in Velichko et al. (2020) by using photoclinometry to account for the change in the photometric longitude component of the disk function.

Using this topographically corrected data, boulder fields in the vicinity of impact craters are identified visually. Polygons encircling boulder fields are mapped and boulder counts are carried out within the delineated polygon. From this, the cumulative fractional area is estimated. During mapping, topographically negative relief features such as small impact craters are avoided. To identify reference rock-free surfaces, we searched around the crater for featureless areas with spatially homogeneous reflectance. The overall reflectance of these areas corresponds to the background reflectance at the regional scale and is less than the reflectance at the crater rim. The details of the selected reference regions are described in section 3.2. We place importance on spatially homogeneous reflectance resulting from flat surfaces and consequently the mapped polygons are 10-50 m wide. We do not use larger boulder field polygons to avoid undulating regional topography. In order to assess the influence of variable unresolved roughness in rock-free surfaces on the NLPRD, we selected multiple reference areas and calculated NLPRD values for each of these areas. For each region, at least one reference area is located away from the crater, outside the blanket and the rays in order to minimize sampling unresolved rocks which might be abundant close to the crater.

In general, where possible, regions with a high slope like within the crater walls or on the crater rim are avoided in order to minimize the influence of topography on the image ratio values. In fact, different incidence angles due to topographic variations from one pixel to another can contribute to a difference in phase ratio values (e.g., Kaydash et al., 2012; Velichko et al., 2022).

# 3. Results

## 3.1 Model results

### 3.1.1 General trends

In figures 4a and 4b the reflectance of simulated rock-rich surfaces of rock abundance at 25% (Figure 3) and 60% are plotted for varying morphologies (Figure 2) as a function of varying phase angle in the emission angle ratio ("configuration 1"). Irrespective of the rock abundance, at very low phase angles the rock-rich surfaces are brighter and their decrease in reflectance is stronger in comparison to the reference rock-free surface.

For rock abundance, 60% between ~50° and 90° phase angles, stark variations in the reflectance slope as well as the absolute values of reflectance in comparison to the rock-free surface is observed. These slope and absolute value variations in the reflectance between the rock-rich surfaces and the rock-free surface are relatively muted at low rock abundance of ~25%. At both rock abundances, subtle differences in reflectance slope between different rock morphologies are identifiable. These differences are investigated in greater detail in subsequent figures.

Contrary to the emission angle ratio (Figure 4a,b), in the incidence angle ratio ("configuration 2"), the relative reflectance between different rock morphologies stays constant across the range of incidence angles (Figure 4c,d).

Figures 5 and 6 present the NLPRD calculated using eq. 6, as a function of rock abundance and reflectance at a particular incidence angle/phase angle, respectively. The different model lines represent scenarios for various possible rock morphologies, i.e., non-abraded, mildly abraded, highly abraded, and non-abraded rocks with tilted top faces.

Overall from figures 5a and b, we find that rock-rich regions do not exhibit uniform photometric behaviour. The abundance and morphology of the individual rocks that constitute the rock-rich regions contribute strongly to the roughness of a rock-rich surface in reference to a rock free surface, as they create slopes on the cm-m scale. In both the geometric configurations, relatively high values of NLPRD can be achieved by modelling non-abraded rocks whereas, for the same height, the tilt of the top of the rock consistently produces low NLPRD values and even much lower than the roughness of the rock-free field.

For example, in both configurations, a rock-rich surface with rocks modelled as non-abraded with top faces tilted at 30° darkens much slower than a rock-free surface. In other words, rock-rich surfaces of morphologies similar to non-abraded with top faces tilted at 30° and highly abraded, are less rough or smoother than a rock-free surface when considered in an incidence ratio configuration as well as an emission ratio configuration. Moreover, the highly abraded morphology that is accompanied by a debris apron/fillet exhibits comparable roughness (NLPRD values) to the non-abraded morphology with the top faces tilted at 30°, indicating that the fillets of rocks play a minor role in characterising the roughness of the rock-rich region.

Figure 6 shows the NLPRD plotted against reflectance at a given phase/incidence angle, and the direction of increasing rock abundance is indicated by the black arrow. In figure 6a, photometrically rougher morphologies, e.g., non-abraded, show a larger spread in reflectance and relatively small variation in NLPRD values. The opposite is observed for highly abraded rocks with a smaller spread in reflectance and greater variation in NLPRD values. In other words, as the morphology of the rocks constituting a boulder field changes, from non-abraded to highly abraded the spread in NLPRD with respect to reflectance increases. Morphologies with tilted top faces at 30˚ and the highly abraded exhibit strongly decreasing NLPRD values accompanied by constant reflectance at low rock abundance. After an inflection point, at high rock abundance, the decrease in NLPRD is accompanied by an increase in reflectance.

Similar to the trends in figure 6a, from figure 6b (configuration 2) it can be seen that various rock morphologies display clear differences in the NLPRD and in the absolute reflectance values. For instance, some rock morphologies like non-abraded display high NLPRD values with a relatively less increase in reflectance. Other morphologies, like highly abraded, have a pronounced increase in reflectance with a decrease in NLPRD. These trends appear to be controlled by the tilting of the rock faces as in figure 6a. As a consequence, the effect of a fillet next to a rock cannot be disentangled from the tilting of the rock top faces. In other words, bright rock fields with a low NLPRD could be explained by rock containing morphologies similar to a highly abraded rock with a fillet or a rock that does not contain a fillet but contains a certain degree of angularity as opposed to a flat-topped rock.

There are additional factors that influence the NLPRD lines such as the choice of scattering properties, the macroscopic roughness parameter ($\bar{\theta}$), the modelled rock size frequency distribution, and the input single scattering albedo values of the rocks and regolith. Their effects on the modelled roughness are discussed in the following sub-sections.

### 3.1.2 Effects of model parameters: Rock SFD

As discussed earlier in section 2.2, in order to assess the effects of rock SFD on the NLPRD lines, we model various rock abundance comprising two large rocks (hereafter addressed as double rock surfaces) as opposed to an exponential distribution (eq. 1).

From figure 7a, at all rock abundances in the emission ratio configuration, the modelled double rock surfaces are rougher (exhibit higher NLPRD values) than surfaces that contain rocks in an exponential SFD for the same rock abundance. This is expected as, for the same rock abundance, the double rock surfaces contain rocks that are larger than the largest rock in the exponential SFD.

For the incidence angle ratio, as seen in figure 8a, a similar trend of double rock surfaces exhibiting higher roughness is preserved. The double rock surfaces cause an increase of ~5x10$^{-3}$ NLPRD for ~24% rock abundance. The increase in NLPRD when considered in an incidence angle ratio (figure 8a) is ~3 times higher than the increase when considered in an emission angle ratio (figure 7a). This further illustrates how resolved shadows dominate the roughness values when considered in the incidence angle ratio.

### 3.1.3. Effect of variable: $\bar{\theta}$

Principally in our modelling approach, we set the $\bar{\theta}$ value for both the regolith and rock to zero in order to only account for roughness and shadowing on the cm-m scale (figure 5, 6). In order to investigate the change in NLPRD when sub-cm scale or roughness in the scale of the microtopography of the rock is included, the $\bar{\theta}$ parameter is given a value of 23.4° (lunar average, Sato et al., 2014) for the rock surfaces. The $\bar{\theta}$ value for the regolith is not modified. Since when calculating the NLPRD, the rock-rich surfaces are always standardized to a rock-free surface, the effect of added $\bar{\theta}$ is negligible for regolith cells, assuming the rock-free surface $\bar{\theta}$ is the same as that of the regolith in between boulders in a rock-rich region. In the emission ratio, NLPRD values are increased by ~3x10$^{-4}$ NLPRD as seen in figure 7b while in the incidence ratio the NLPRD values are increased by ~9x10$^{-4}$ NLPRD in figure 8b.

In summary, for both the emission ratio as well as the incidence ratio, we find that introduction of sub-cm scale roughness via the Hapke macroscopic roughness parameter $\bar{\theta}$ causes an increase in the NLPRD values.

### 3.1.4. Effects of model parameters: *c*

In our default modelling approach, we model rock and regolith surfaces as backscattering – likened to a case where the rock is covered by a layer of dust that is assumed to behave as the regolith. As discussed earlier, most rock surfaces are roughly isotropic scatterers (Hapke and Van Horn, 1963). In order to model a case where the rock surface is uncovered by dust and patina, the rock is modelled as an isotopic scatterer, i.e., $p(\alpha) = 1$ and the variations in NLPRD as a result of isotropic scattering of rocks is shown in figure 7c and 8c for the emission angle ratio and incidence angle ratio, respectively.

In the case of the emission ratio and incidence ratio, the effects of changing the scattering behaviour of the rock from a backscattering formalism to an isotropic scattering formalism result in a mild (~0.7x10$^{-3}$) decrease in NLPRD with increasing rock abundance.

### 3.1.5. Effects of model parameters: *ω*

Figures 7d and 8d illustrate the variation in the NLPRD values that occur as a result of a decrease in the value of $\omega_{rock}$. In the emission ratio (figure 7d) a small change in the NLPRD behaviour is observed when the values of $\omega_{rock}$ are set equal to the $\omega_{regolith}$. At a rock abundance of ~24%, the NLPRD of a rock-rich surface of "dark" rocks is ~0.5x10$^{-3}$ NLPRD higher than that of a rock-rich surface of bright rocks.

For the incidence ratio (figure 8d) the trend of an increase in NLPRD with increasing rock abundance is preserved when the albedo of rocks ($\omega_{rock}$) is reduced to match that of the regolith ($\omega_{regolith}$). However, the increase in NLPRD is twice than that observed in the emission ratio. In summary, in both the image ratios, a rock-rich surface consisting of "darker" rocks is photometrically rougher than when the rock-rich surface is comprised of bright rocks.

### 3.2 Comparison with NAC data of rock-rich areas

## 3.2.1 Emission Angle Ratio – Configuration 1(Young crater on the inner flank of Hertzsprung S)

In this section, we briefly describe the spatial context of our boulder fields and the reference regions. We then present the variance in the NLPRD exhibited by the boulder fields especially when considered with multiple rock-free reference regions. Finally, an outline of the comparison between NAC-derived NLPRD and model-derived NLPRD is presented.

We analyze the NLPRD for 23 different boulder fields selected in the vicinity of the unnamed crater on the inner flank of Hertzsprung S. These boulder fields are sampled across regions on the ejecta blanket up until ~3 crater radii. The location of these boulder fields (marked in numbered gray circles) in association with radial distance from the crater is shown in figure 9. Additionally, we select three different rock-free reference regions based on their location on ($M_1$, $M_3$) or away from ($M_2$) the highly reflective ejecta blanket. The rock-free reference regions are shown in maroon crosses also in figure 9. A zoom-in of the sampled boulder fields and the reference rock-free regions is presented in figure 10, in order to provide spatial context to the corresponding NLPRD for every boulder field.

The use of topographically corrected data leads to an expected decrease of the NLPRD values relative to non-corrected data from a minimum of ~$0.016 \times 10^{-3}$ NLPRD to a maximum of $1.6 \times 10^{-3}$ NLPRD. Figure 11 shows the NLPRD of the sampled boulder fields as a function of the rock abundance expressed as cumulative fractional area percentage. As mentioned earlier we calculate the NLPRD of the boulder fields with respect to three rock-free reference regions. In figure 11 we choose to only plot the NLPRD of the boulder fields that are calculated with the roughest (reference region with the lowest phase ratio- denoted $M_2$) and the smoothest (reference region with the highest phase ratio - denoted $M_3$) reference regions, i.e., only the lowest and the highest NLPRD of the boulder fields are shown. The third reference region exhibited intermediate phase ratio values and produced intermediate NLPRD values of the boulder fields. NLPRD calculated using the third reference region is ignored in the plot to maintain clarity. It is seen that boulder fields exhibit varying roughness characteristics despite being visually indistinguishable (Figure 10). The boulder fields that exhibit the lowest values of NLPRD are predominantly situated in the eastern region of the ejecta blanket of the young crater.

Figure 11 also shows the model-derived NLPRD lines for various rock morphologies. The model-derived NLPRD lines for various rock morphologies are unable to reproduce exactly the roughness exhibited by all the boulder fields. Only the smoothest boulder fields, i.e., B0, B5, and B9, plot in proximity to the non-abraded morphology. The possible reasons for this mismatch are discussed in section 4.

### 3.2.2 Incidence angle Ratio - configuration 2 (North Ray crater)

The second study area is near the Apollo 16 landing site: North Ray crater on the highlands. Boulder fields were sampled across the ejecta blanket. The locations of the sampled boulder fields (in red) in association with radial distance from the crater center are shown in figure 12. The sampled rock-free reference regions are also shown in figure 12 by white crosses. The undulating regional topography, namely Stone Mountain and a ravine to the east, and Kiva crater to the west, strongly restricted the sampling of boulder fields across the ejecta blanket.

Four different rock-free reference surfaces are sampled in the west, south, and north of the crater. They are referred to henceforward as $W_1$, $W_2$, and $S_1$, $N_1$ where W, S, N refer to their directional location in reference to North ray crater.

The $W_1$ surface is located on the highly reflective ejecta blanket while $W_2$ is away from the bright rays. The $S_1$ surface is on the highly reflective ejecta blanket on the south east flank (located on the North Ray ridge) while $N_1$ is chosen north of the crater, to ensure representativeness in relation to the sampled boulder fields. A zoom-in of the sampled boulder fields and the four reference rock-free regions is presented in figure 13.

From the phase ratio values of the rock-free reference regions alone, we find that the reference region $W_2$ is the smoothest with the highest values of the phase ratio while the roughness of the remaining reference regions is quite similar with reference $N_1$ being the roughest, i.e., lowest phase ratio value.

From figure 14 we find that the rock fields exhibit a wide range of NLPRD values irrespective of the different reference regions. Few boulder fields exhibit negative NLPRD values, indicating that their surface roughness is smoother than the reference rock-free regions. This essentially implies that the photometric roughness is not dominated by the large-scale shadowing by fully resolved rocks but possibly by the roughness on the sub-cm scale of the reference rock free field- as the smoothest reference region W2 produces the least number of boulder fields with below zero NLPRD.

A majority of the boulder fields that exhibit low NLPRD values (implying surface roughness comparable to rock-free reference surface) are situated on the south east of North Ray crater (as seen in figure 14). This trend is preserved when rock fields are normalised to all the four-reference rock-free surfaces, as shown in figure 14. It is to be noted that even if care was taken to avoid well-resolved topographic variations not due to rocks, we cannot completely exclude the presence of subdued reliefs, like craters or ridge-like features few pixels in size in boulder fields 2, 3, and 13. Such subdued reliefs would increase the NLPRD for such areas. Nevertheless, this effect is negligible because in the boulder field 6 that has large roughness, there are no visually well-resolved relief features apart from the resolved boulders.

From the comparison between model derived and NAC derived NLPRD, we find that the spread of the model derived NLPRD for varying rock morphologies is far exceeded by the spread of the NLPRD values of the NAC resolved boulder fields.

## 4. Discussion

### 4.1. Default modelling approach

This section discusses the NLPRD values and the regional patterns exhibited by boulder fields in the ejecta blanket of the two impact craters – unnamed crater at Hertzsprung S and North Ray. Additionally, this section also discusses the results of the comparison between the NLPRD of the NAC-resolved boulder fields with the default model-derived NLPRD. The default modelling approach refers to the scenario where the artificial rock-rich surface follows an exponential rock SFD (eq 1), where both the rocks and the regolith are principally backscattering and where the SSA of the rock is set higher than the SSA of the regolith. Additionally, for the default approach the $\bar{\theta}$ parameter (correction for macroscopic roughness) is ignored.

(i) **Unnamed crater at Hertzsprung S**

Overall, from figure 11, we find that majority (with the exception of boulder field 13) of the boulder fields situated on the west are smoother i.e., show lower NLPRD values when compared to the other sampled boulder fields. This finding holds true when evaluated with rock-free reference surface on and off the highly reflective ejecta blanket. The trend of asymmetric roughness characteristics is broadly in agreement with the results from Velichko et al. (2022) indicating that the unnamed crater exhibits ejecta asymmetry attributed to an oblique impact. Their study of the pixel-by-pixel variation of the phase slope showed that the ejecta blanket to the west of the crater is smoother, i.e., exhibits a shallower phase curve which implies that the impact was westward. The smoothness of the regolith of the west has been discussed to be due to the formation of the DHRZ (distal highly reflective zone) by high-velocity vapor jets (Speyerer et al., 2016) that in turn destroys the porosity of the upper surface of the regolith and thereby reduces the surface roughness. We also note that in the study of Velichko et al.,(2020) vapour wind shadows have been associated with boulders. It is possible that the high-velocity jetted vapor contributes to modification or smoothening of the dust that is present on the boulders. However, this process is inconsistent with the sequence of processes during the impact event because the high-velocity jets occur early on in the impact event before the excavation stage (Birkhoff et al., 1948; Walsh et al., 1953; Johnson et al., 2014). However, in order for the vapour jets to modify the surface of boulders, the boulders must already be in place prior to the vapour jetting.

Alternatively, as the smooth boulder fields lie on the west, i.e., the downrange direction of the oblique impact, it could also be possible that the micro-texture of these boulders were preferentially shock modified in comparison to the boulders in the uprange zone. This could represent a type of asymmetric shock deformation known to occur during oblique impact process (Pierazzo and Melosh, 2000b; Kenkmann et al., 2014). These differently shocked ejecta boulders in the downrange could also respond differently to space weathering after emplacement (Pieters and Noble., 2016; Patzek and Rüsch, 2022) when compared to boulders in the forbidden/uprange region. Moreover, we note that the impact crater is situated on a ridge within the Hertzsprung S basin, and this kms-scale irregular topography could have contributed to the asymmetric ejecta distribution. Overall our results suggest that the signatures of ejecta asymmetry due to oblique impacts are not restricted to the microtopography of the regolith but also to the surface (microtopography) of the rocks that constitute the ejecta. Another scenario to explain the observed spatial pattern of photometrically smooth and rough boulders being located on either side of the impact crater is to evoke target heterogeneities, i.e., petrological or rheological heterogeneities in the excavated bedrock that pre-date the impact event.

Apart from general trends of ejecta asymmetry, we find that boulder field 4, which is essentially a "single" boulder that was emplaced by rolling and/or bouncing (similar to those observed in the study of Kumar et al. (2016)), exhibits the highest value of NLPRD. A high roughness could be explained by an irregular topography, e.g., the nearby regolith was likely disturbed by the moving boulder. However, the area affected by the trail is excluded from the polygon. Therefore, we cannot exclude a contribution from the rock surface. We hypothesize that this boulder constitutes the latest event in the impact process that formed the unnamed crater. We also find that no vapour wind shadow is associated with this rolling boulder indicating that there might have been no textural modification due to the jetted vapour and thereby this boulder appears photometrically rougher. This high roughness could therefore be representative of the pristine unmodified surface of the ejecta boulder.

With respect to the fit between the NAC derived NLPRD and the model derived NLPRD (in the default approach-as described earlier) we find that only the smoothest boulder fields, i.e., boulder fields 0, 5, and 9 exhibit NLPRD values in proximity to the "roughest" modelled morphology i.e., Non-Abraded. This essentially implies that the roughness of the NAC-resolved boulder fields is not fully explained by the cm-m scale roughness and there exists a sub-resolution roughness that needs to be considered and investigated. This is elaborated in section 4.2.

### (ii) North Ray crater

Similar to the case of the unnamed crater at Hertzsprung S, there is a correlation between our findings at North Ray crater and previously reported ejecta blanket heterogeneity. The ejecta blanket heterogeneity represented by the observed spatial clustering of boulders with low or high NLPRD values on either side of North Ray crater is spatially consistent with the large-scale photometric heterogeneity reported in the study of Watkins et al. (2021). In particular, they find variation in the *b* parameter of the phase function for regions on the west and southeast of North Ray. The *b* parameter is slightly lower in the southeast region of North ray - thereby indicating slightly stronger backscattering properties. Watkins et al. (2021) speculate that the changes in the *b* values could be indicative of changes in the physical properties of the ejecta. The southeast region of North Ray is also discussed in the study by Gao et al. (2022) which uses the backscatter intensity (BSI) data from the Mini-RF. The BSI indicates variation in decimetre scale roughness – with the south eastern region being rougher than the west. This contradiction in surface roughness patterns between our findings and those of Gao et al. (2022) could be attributed to the fact that we are indeed comparing roughness at two distinct scales. The same scenarios evoked for the unnamed crater are valid to explain the observations at North Ray crater. Because the spatial clustering of NLPRD values is measured at both craters, the cause is more likely associated to the impact crater process rather than pre-existing petrological or rheological heterogeneities in the excavated bedrock.

Boulder field 11 shows consistently the lowest NLPRD values irrespective of the sampled reference region. Also, boulder fields 9 and 10 show low NLPRD values comparable to boulder field 11. Upon a closer look at these boulder fields (see figure 13), we see that they contain a few diffused patterns that do not cast a strong shadow as sharply resolved rocks do. We interpret these diffused patterns to be due to mound-like abraded rocks with a fillet similar to the highly abraded morphology. Moreover, the upper layer of the filleted material could be relatively younger than the rock-free unit – and hence have a higher reflectance and albedo (e.g., Lucey et al., 1995; Sasaki et al., 2001) that reduces the photometric roughness by reducing the shadows between particles through multiple scattering (Buratti & Veverka, 1985). This has also been documented in Hapke (1984,1993) where particles with higher albedo negate the shadows and reduce the photometric roughness.

In general, the presence of unresolved rocks or cobbles that are considered as unresolved roughness in the intermediate scale (~1mm-0.5m) could explain the elevated roughness observed in the boulder fields. In NAC images we see multiple instances of bright pixels sometimes accompanied by an adjacent darker pixel which we tentatively interpret to be rocks of diameter lesser than the image resolution i.e., rock fragments or cobbles well below 1m in diameter. However, due to limitations in the currently available data resolution, along with the known artefacts in the LRO NAC data such as the echo effect (Hamm et al., 2016), further interpretation of these bright and dark pixels is limited. Nonetheless, it is unlikely that unresolved rocks are responsible for the measured elevated roughness because model NLPRD values do not increase significantly with increasing rock abundance (Figure 15).

As we move to the roughest group of boulder fields, i.e., the boulder fields locate on the west, the fit between model-derived NLPRD and the NAC-derived NLPRD becomes poor. Effectively this suggests that our model surfaces do not contain roughness at a particular size scale. In our model synthetic surfaces, we model rocks and the debris aprons that contribute to slopes in the cm-m scale. However, we do not model any slope or topography exclusively on the surface of rocks – below the cm scale. In order to explore this possibility, we invoke variations in the Hapke macroscopic roughness parameter as discussed in the next section.

Also, when we compare the synthetic surfaces generated using the exponential formalism (figure 3), for the same CFA, the NAC-resolved boulder fields do not appear to have a similar rock size distribution. And hence we test the effects of these parameters in the next section.

### 4.2. Modelling approach utilizing additional parameters

In this section, we discuss the variations in modelled roughness that could explain the data point, by modification of existing parameters (rock SFD, SSA) and/or including new parameters ($\bar{\theta}$).

By changing the rock SFD, effectively we alter slopes/topography in the cm-m scale. The persisting mismatch between the NAC-derived NLPRD and model-derived NLPRD could imply the key role of surface topography and thereby roughness at a scale that is not modelled, i.e., the sub cm scale roughness. From previous studies (Hapke 1984, Shepard and Helfenstein, 2007) it is understood that $\bar{\theta}$ is a correction term introduced to account for the slopes/roughness that occurs at all scales up until the scale of the pixel resolution. Hence, in an attempt to simulate sub-cm scale roughness, we introduce $\bar{\theta}$ values of the rock surface. We find that by increasing the macroscopic roughness value for rock surfaces, larger NLPRD are reached. Our results suggest that by altering the rock SFD and by accounting for the sub-cm scale roughness of the rock surfaces, the model will be able to explain the roughness of these boulder fields. However, we refrain from reporting a best fit $\bar{\theta}$ value for two reasons. First, constraining the Hapke parameters, especially $\bar{\theta}$, in a non-ambiguous way requires more data points along the phase curve, in particular at high phase angles (~80°) (e.g., Verbiscer and Veverka, 1989; Pilorget et al., 2016; Li et al., 2021; Jiang et al., 2021; Golish et al., 2021). Second, geologically plausible $\bar{\theta}$ values for the sub-cm scale of lunar (and terrestrial) rocks to be used as input range in a model are insufficiently known.

From figures 7d and 8d, we also find that the effect of decreasing single scattering albedo ($\omega$) results in similar increase in NLPRD as when the correction for macroscopic roughness ($\bar{\theta}$) is included, thereby indicating similar contribution of these two parameters to the darkening of the rock-rich surface (in comparison to the rock- free reference regions) in the phase angle range considered.

It is understood from literature that the photometric parameters that describe the micro-texture and composition of the regolith and rock surfaces are highly interconnected via the petrology (e.g., Hapke, 1993; Shepard & Campbell, 1998; Shepard & Helfenstein, 2007; Pilorget et al., 2016; Sun et al., 2022) thereby warranting a detailed investigation of the photometric properties of unpolished rock surfaces and the evolution of these properties in response to space weathering and thermal fatigue.

Finally, when modelling fully resolved boulder fields we assume that all rocks comprising the boulder field have the same morphology. This assumption is justified for low rock abundance, boulder fields of a small area, and in particular sub-pixel boulder fields, making the NLPRD a useful parameter to assess the morphology of sub-pixel (unresolved) boulder fields. However, in the case of large resolved boulder fields with a higher rock abundance, our assumption that all rocks modelled belong to the same family may not be appropriate, as in the case of boulder field 8 (figure 13) where different expressions of rock abrasion are visible. For that site, mound-like diffused and abraded rocks are in the same boulder field polygon as rocks of no identifiable fillets.

## 5. Conclusions

In this study, the influence of rocks on the photometry of the lunar surface for different geometric configuration is investigated by utilising a phase ratio – which is an effective way to gauge surface roughness information using limited imagery. The surface roughness is parameterized by the Normalised Logarithmic Phase Ratio Difference (NLPRD). The NLPRD measures how rapidly a rock-rich surface decreases in reflectance over a range of phase angles in comparison to a surface that does not host rocks.

As the NLPRD is normalised to a rock-free reference surface, provided the roughness of the regolith in the rock-free reference region is similar to that considered within the boulder field, the resulting roughness values represent roughness exclusive to the rock surface and shape. In our modelling approach, we explore the variability of topography in the cm-m scale through the rock size frequency distribution and rock morphologies. In the second set of modelling iterations, we investigate the influence of roughness/slopes that lie far below the scale previously considered, i.e., in the form of the Hapke macroscopic roughness which is dominated by the strongest slopes at the smallest scale. Quantifying the effect of changing the sub-resolution particle physical properties and composition is attempted by modifying the scattering parameters and the single scattering albedo as they are known to be linked to the aforementioned properties (Hapke 1993; Helfenstein et al., 1988; Cord et al., 2003; Sato et al., 2014). However, we wish to remark that in this second modeling approach, we choose to use the Hapke macroscopic roughness parameter (via the shadowing function) to account for the sub-resolution photometric roughness despite extensive discussion in literature regarding its ambiguities, in particular, the assumption of a gaussian distribution of slopes on planetary surfaces that is likely unrealistic (Shkuratov et al., 2012; Davidsson et al., 2015; Labarre et al., 2017; Hapke, 2013; Hasselmann et al., 2021). Moreover, alternative models that account for the photometric effects of surface roughness exist (eg: van Ginneken et al., 1998; Lumme and Bowell, 1981; Shkuratov and Helfenstein, 2001; Shiltz and Bachmann,2023). Future understanding will benefit from a detailed characterization of the microtopography of rock surfaces along with utilizing other existing photometric models to describe surface roughness.

The modelled results have been compared to data points from the LROC NAC images. Below we summarise the key findings of this study:

- Overall, we find that rock size frequency distribution and rock morphology i.e., relief that creates topography in the cm-m scale, have significant effects on the normalised photometric roughness whereas the influence of the Hapke parameters i.e., rock single scattering albedo, scattering behaviour, and macroscopic roughness is minor but not negligible.

- Different rock morphologies (shapes) can create the same photometric roughness and hence it is not possible to ascertain rock morphology based on photometry alone. Some rock shapes can result in no photometric roughness at all relative to a rock-free surface. Therefore, contrary to previous understanding, rock-rich surfaces are not strictly photometrically rougher than rock-free surfaces. This is valid for rock-rich surfaces both spatially resolved and at the sub-pixel level.

- Ideally, additional insight on rock morphology can be gained by using a combination of incidence (image ratio where the two images are acquired from the nadir at varying incidence angles) and emission (image ratio where the two images are acquired at same incidence angles) ratios image for the same region.

Regarding currently available data:

- Boulder fields that are visually and photo-geologically indistinguishable in LROC NAC data have different photometric roughness characteristics in image ratio. Boulder fields, even from the same impact crater, can be both photometrically smoother or rougher than rock-free units.

- Variations in the photometric surface roughness of boulder fields are likely indicative of ejecta asymmetry that is directly linked to oblique impact crater conditions. The surface of ejecta blocks is not necessarily pristine and might be modified during the impact crater excavation process. This suggests that the signature of ejecta asymmetry is not limited to changes in the surface properties of regolith but also extends to the surface properties of ejecta blocks. In general, our results imply that the rock physical properties at the start of the surface exposure period are a function of petrology as well as the (shock) effects imparted upon ejecta rock formation and excavation.

## 6. Acknowledgements

RMM and OR are supported by a Sofja Kovalevskaja award of the Alexander von Humboldt foundation. Two anonymous referees are acknowledged for their constructive comments. Discussion of the geology of the Apollo 16 landing site with Wajiha Iqbal is acknowledged. We thank the LROC team for the acquisition of the images.

## 7. Data availability

LROC/NAC image data are available at http://wms.lroc.asu.edu/lroc/search.

## 8. References:

Adams, J., & McCord, T. (1971). Optical properties of mineral separates, glass, and anorthositic fragments from Apollo mare samples. Lunar and Planetary Science Conference Proceedings, 2, 2183.

Akimov, L. A. (1988a). Light reflection by the Moon. I. Kinematika i Fizika Nebesnykh Tel, 4, 3-10.

Akimov, L. A. (1988b). Light reflection by the moon. II. Kinematika i Fizika Nebesnykh Tel, 4, 10-16.

Akimov, L. A., Velikodskij, Y. I., & Korokhin, V. V. (1999). Dependence of lunar highland brightness on photometric latitude. Kinematics and Physics of Celestial Bodies, 15(4), 232-236.

Akimov, L. A., Velikodskiy, Y. I., & Korokhin, V. V. (2000). Dependence of latitude brightness distribution over the Lunar disk on albedo and relief. Kinematika i Fizika Nebesnykh Tel, 16(2), 181-187.

Bandfield, J. L., Ghent, R. R., Vasavada, A. R., Paige, D. A., Lawrence, S. J., & Robinson, M. S. (2011). Lunar surface rock abundance and regolith fines temperatures derived from LRO Diviner Radiometer data. Journal of Geophysical Research E: Planets, 116(12). https://doi.org/10.1029/2011JE003866

Basilevsky, A. T., Head, J. W., & Horz, F. (2013). Survival times of meter-sized boulders on the surface of the Moon. Planetary and Space Science, 89, 118–126. https://doi.org/10.1016/j.pss.2013.07.011

Belgacem, I., Schmidt, F. and Jonniaux, G. (2021). Regional Study of Ganymede's photometry. Icarus, 369, p. 114631. https://doi.org/10.1016/j.icarus.2021.114631.

Bickel, V.T., Aaron, J., Manconi, A. et al. Impacts drive lunar rockfalls over billions of years. Nat Commun 11, 2862 (2020). https://doi.org/10.1038/s41467-020-16653-3

Birkhoff, G., MacDougall, D.P., Pugh, E.M., Taylor, Geoffrey, S., 1948. Explosives with Lined Cavities. J. Appl. Phys. 19, 563–582. http://dx.doi.org/10.1063/1.1698173

Buratti, B. J., & Veverka, J. (1985). Photometry of rough planetary surfaces: The role of multiple scattering. Icarus, 64(2), 320–328. https://doi.org/10.1016/0019-1035(85)90094-6

Clark, B. E., Hapke, B., Pieters, C., & Britt, D. (2002). Asteroid space weathering and regolith evolution. Asteroids III, 585, 90086-2

Clegg, R. N., B. L. Jolliff, M. S. Robinson, B. W. Hapke, and J. B. Plescia. Effects of rocket exhaust on lunar soil reflectance properties. Icarus (2014), 227, 176-194. https://doi.org/10.1016/j.icarus.2013.09.013

Clegg-Watkins, R. N., B. L. Jolliff, A. Boyd, M. S. Robinson, R. Wagner, J. D. Stopar, and J. B. Plescia (2016). Photometric characterization of the Chang'e-3 landing site using LROC NAC images, Icarus, 273, 84-95. https://doi.org/10.1016/j.icarus.2015.12.010

Cord, A., Baratoux, D., Mangold, N., Martin, P., Pinet, P., Greeley, R., Costard, F., Masson, P., Foing, B., & Neukum, G. (2007). Surface roughness and geological mapping at subhectometer scale from the High Resolution Stereo Camera onboard Mars Express. Icarus, 191(1), 38–51. https://doi.org/10.1016/j.icarus.2007.04.029

Davidsson, B. J., Rickman, H., Bandfield, J. L., Groussin, O., Gutiérrez, P. J., Wilska, M., Capria, M.T., Emery, J. P., Helbert, J., Jorda, L., Maturilli, A., Mueller, T. G., 2015. Interpretation of thermal emission. I. The effect of roughness for spatially resolved atmosphereless bodies. Icarus 252, 1–21. https://doi.org/10.1016/ j.icarus.2014.12.029.

De Angelis, S., Manzari, P., De Sanctis, M. C., Altieri, F., Carli, C., & Agrosì, G. (2017). Application of spectral linear mixing to rock slabs analyses at various scales using Ma_Miss BreadBoard instrument. Planetary and Space Science, 144, 1–15. https://doi.org/10.1016/j.pss.2017.06.005

Demidov, N. E., & Basilevsky, A. T. (2014). Height-to-diameter ratios of moon rocks from analysis of Lunokhod-1 and -2 and Apollo 11–17 panoramas and LROC NAC images. Solar System Research, 48(5), 324–329. https://doi.org/10.1134/S0038094614050013

Domingue, D., Palmer, E., Gaskell, R., & Staid, M. (2018). Characterization of lunar surface within Tsiolkovsky crater: Photometric properties. Icarus, 312, 61–99. https://doi.org/10.1016/j.icarus.2018.02.034

Fortezzo, C. M., Spudis, P. D., & Harrel, S. L. (2020). Release of the digital unified global geologic map of the Moon at 1: 5,000,000-Scale. 51st Lunar and Planetary Science Conference.

Gao, Y., Liu, X., Hou, W., Han, Y., Zhang, H., & Wang, R. (2022). Dielectric Constant Estimation for Apollo 16 site using Mini-RF SAR data. 53rd Lunar and Planetary Science Conference.

Ghent, R.R. et al. (2014) "Constraints on the recent rate of lunar ejecta breakdown and implications for crater ages," Geology, 42(12), pp. 1059–1062. https://doi.org/10.1130/g35926.1.

Golish, D. R., Li, J., Clark, B. E., DellaGiustina, D. N., Zou, X., Rizos, J. L., . . . Lauretta, D. S. (2021). Regional Photometric Modeling of asteroid (101955) Bennu. The Planetary Science Journal, 2(4), 124. https://doi.org/10.3847/psj/abfd3c

Grumpe, A., Belkhir, F., & Wöhler, C. (2014). Construction of lunar DEMs based on reflectance modelling. Advances in Space Research, 53(12), 1735–1767. https://doi.org/10.1016/j.asr.2013.09.036

Hamm, D. C., Tschimmel, M., Brylow, S. M., Mahanti, P., et al., 2016, Flight calibration of the LROC narrow angle camera, Space Sci. Rev., 200, 431–473, https://doi.org/10.1007/s11214-015-0201-8

Hapke, B. (1981). Bidirectional reflectance spectroscopy: 1. Theory. Journal of Geophysical Research: Solid Earth, 86(B4), 3039–3054. https://doi.org/10.1029/jb086ib04p03039

Hapke, B. (1984). Bidirectional reflectance spectroscopy. 3. Correction for macroscopic roughness. Icarus, 59(1), 41–59. https://doi.org/10.1016/0019-1035(84)90054-X

Hapke, B. (1993). Theory of reflectance and emittance spectroscopy. Topics in Remote Sensing.

Hapke, B. (1999). Scattering and diffraction of light by particles in planetary regoliths. Journal of Quantitative Spectroscopy and Radiative Transfer, 61(5), 565–581. https://doi.org/10.1016/S0022-4073(98)00042-9

Hapke, B. (2002). Bidirectional Reflectance Spectroscopy: 5. The Coherent Backscatter Opposition Effect and Anisotropic Scattering. Icarus, 157(2), 523–534. https://doi.org/10.1006/icar.2002.6853

Hapke, B. (2008). Bidirectional reflectance spectroscopy. 6. Effects of porosity. Icarus, 195(2), 918–926. https://doi.org/10.1016/j.icarus.2008.01.003

Hapke, B., 2013. Comment on "A critical assessment of the Hapke photometric model. Journal of Quantitative Spectroscopy and Radiative Transfer 116, 184–190. https://doi.org/10.1016/j.jqsrt.2012.11.002.

Hapke, B., & van Horn, H. (1963). Photometric studies of complex surfaces, with applications to the Moon. Journal of Geophysical Research, 68(15), 4545–4570. https://doi.org/10.1029/JZ068I015P04545

Hasselmann, P. H., Fornasier, S., Barucci, M. A., Praet, A., Clark, B. E., Li, J. Y., Golish, D. R., DellaGiustina, D. N., Deshapriya, J. D. P., Zou, X. D., Daly, M. G., Barnouin, O. S., Simon, A. A., & Lauretta, D. S. (2021). Modeling optical roughness and first-order scattering processes from OSIRIS-REx color images of the rough surface of asteroid (101955) Bennu. Icarus, 357, 114106. https://doi.org/10.1016/j.icarus.2020.114106

Hawke, B. R., Lucey, P. G., Taylor, G. J., Bell, J. F., Peterson, C. A., Blewett, D. T., . . . Spudis, P. D. (1991). Remote Sensing Studies of the orientale region of the Moon: A pre-Galileo View. Geophysical Research Letters, 18(11), 2141-2144. https://doi.org/10.1029/91gl02667

Helfenstein, P., & Shepard, M. K. (1999). Submillimeter-Scale Topography of the Lunar Regolith. Icarus, 141(1), 107–131. https://doi.org/10.1006/icar.1999.6160

Helfenstein, P., Veverka, J., & Thomas, P. C. (1988). Uranus satellites: Hapke parameters from Voyager disk-integrated photometry. Icarus, 74(2), 231-239. https://doi.org/10.1016/0019-1035(88)90039-5

Helfenstein, P., & Veverka, J. (1987). Photometric properties of lunar terrains derived from Hapke's equation. Icarus, 72(2), 342-357. https://doi.org/10.1016/0019-1035(87)90179-5

Hörz, F., Basilevsky, A. T., Head, J. W., & Cintala, M. J. (2020). Erosion of lunar surface rocks by impact processes: A synthesis. Planetary and Space Science, 194, 105105. https://doi.org/10.1016/j.pss.2020.105105

Hörz, F., Schneider, E., Gault, D.E. et al. (1975). Catastrophic rupture of lunar rocks: A Monte Carlo simulation. The Moon, 13, 235–258 https://doi.org/10.1007/BF00567517

Jiang, T., Hu, X., Zhang, H., Ma, P., Li, C., Ren, X., Lin, H., 2021. In situ lunar phase curves measured by Chang'e-4 in the von Kármán Crater, South Pole-Aitken Basin. Astronomy & Astrophysics 646. https://doi.org/10.1051/0004-6361/202039252.

Jin, W., Zhang, H., Yuan, Y., Yang, Y., Shkuratov, Y. G., Lucey, P. G., Kaydash, V. G., Zhu, M. H., Xue, B., Di, K., Xu, B., Wan, W., Xiao, L., & Wang, Z. (2015). In situ optical measurements of Chang'E-3 landing site in Mare Imbrium: 2. Photometric properties of the regolith. Geophysical Research Letters, 42(20), 8312–8319. https://doi.org/10.1002/2015GL065789

Johnson, B. C., Bowling, T. J., Melosh, H. J., 2014, Jetting during vertical impacts of spherical projectiles, Icarus, 238, 13-22. https://doi.org/10.1016/j.icarus.2014.05.003

Johnson, J. R., Grundy, W. M., & Shepard, M. K. (2004). Visible/near-infrared spectrogoniometric observations and modeling of dust-coated rocks. Icarus, 171(2), 546–556.

https://doi.org/10.1016/j.icarus.2004.05.013

Jolliff, B. L., Schonwald, A. R., & Watkins, R. N. (2020). Assessment of Anorthosite on the Moon Using LROC Narrow Angle Camera Photometry: Correlations Between Single Scattering Albedo, Composition, and Optical Maturity. 51st Annual Lunar and Planetary Science Conference

Kaydash, V., & Shkuratov, Y., (2012). Structural disturbances of the lunar surface caused by spacecraft. Solar System Research, 46(2), 108–118. https://doi.org/10.1134/S0038094612020050

Kaydash, V., Shkuratov, Y., & Videen, G. (2012). Phase-ratio imagery as a planetary remote-sensing tool. Journal of Quantitative Spectroscopy and Radiative Transfer, 113(18), 2601–2607. https://doi.org/10.1016/j.jqsrt.2012.03.020

Kaydash, V., Shkuratov, Y., Korokhin, V., & Videen, G. (2011). Photometric anomalies in the Apollo landing sites as seen from the Lunar Reconnaissance Orbiter. Icarus, 211(1), 89–96. https://doi.org/10.1016/j.icarus.2010.08.024

Kenkmann, T., Poelchau, M. H., & Wulf, G. (2014). Structural geology of impact craters. Journal of Structural Geology, 62, 156-182. https://doi.org/10.1016/j.jsg.2014.01.015

Kreslavsky, M. A., & Shkuratov, Y. (2003). Photometric anomalies of the lunar surface: Results from Clementine data. Journal of Geophysical Research: Planets, 108(E3). https://doi.org/10.1029/2002JE001937

Krishna, N., & Kumar, P. S. (2016). Impact spallation processes on the Moon: A case study from the size and shape analysis of ejecta boulders and secondary craters of Censorinus crater. Icarus, 264, 274–299. https://doi.org/10.1016/j.icarus.2015.09.033

Kumar, P.S, Sruthi, U., Krishna, N., Lakshmi, K. J. P., Menon, R., Amitabh, , Gopala Krishna, B., Kring, D. A., Head, J. W., Goswami, J. N., et al. (2016), Recent shallow moonquake and impact-triggered boulder falls on the Moon: New insights from the Schrödinger basin, *J. Geophys. Res. Planets*, 121, 147– 179, doi:10.1002/2015JE004850.

Labarre, S., Ferrari, C., & Jacquemoud, S. (2017). Surface roughness retrieval by inversion of the Hapke model: A multiscale approach. Icarus, 290, 63–80. https://doi.org/10.1016/j.icarus.2017.02.030

Li, J., Zou, X., Golish, D. R., Clark, B. E., Ferrone, S., Fornasier, S., . . . Lauretta, D. S. (2021). Spectrophotometric modeling and mapping of (101955) Bennu. The Planetary Science Journal, 2(3), 117. https://doi.org/10.3847/psj/abfd2d

Li, Y., & Wu, B. (2018). Analysis of rock abundance on lunar surface from orbital and descent images using automatic rock detection. Journal of Geophysical Research: Planets, 123, 1061– 1088.

https://doi.org/10.1029/2017JE005496

Lin, H., Yang, Y., Lin, Y., Liu, Y., Wei, Y., Li, S., Hu, S., Yang, W., Wan, W., Xu, R., He, Z., Liu, X., Xing, Y., Yu, C., & Zou, Y. (2020). Photometric properties of lunar regolith revealed by the Yutu-2 rover. Astronomy & Astrophysics, 638, A35. https://doi.org/10.1051/0004-6361/202037859

Lucey, P. G., Taylor, G. J., & Malaret, E. (1995). Abundance and distribution of iron on the Moon. Science, 268(5214), 1150-1153. https://doi.org/10.1126/science.268.5214.1150

Lumme, K., Bowell, E., 1981. Radiative transfer in the surfaces of atmosphereless bodies. I-Theory. II-Interpretation of phase curves. The Astronomical Journal 86, 1694–1721.

McGuire, A. F., & Hapke, B. W. (1995). An Experimental Study of Light Scattering by Large, Irregular Particles. Icarus, 113(1), 134–155. https://doi.org/10.1006/icar.1995.1012

Muehlberger, W. R., Batson, R. M., Boudette, E. L., Duke, C. M., Eggleton, R. E., Elston, D. P., ... & Young, J. W. (1972). Preliminary geologic investigation of the Apollo 16 landing site.

Noguchi, T., Tsuchiyama, A., Hirata, N., Demura, H., Nakamura, R., & Miyamoto, H. et al. (2010). Surface morphological features of boulders on Asteroid 25143 Itokawa. Icarus, 206(1), 319-326. https://doi.org/10.1016/j.icarus.2009.09.002

Patzek, M., & Rüsch, O. (2022). Experimentally induced thermal fatigue on lunar and eucrite meteorites—influence of the mineralogy on rock breakdown. Journal of Geophysical Research: Planets, 127, e2022JE007306. https://doi.org/10.1029/2022JE007306

Pierazzo, E., Melosh, H. J., 2000b. Melt Production in Oblique Impacts. Icarus 145 (1), 252–261. https://doi.org/10.1006/icar.1999.6332.

Pieters, C. M., & Noble, S. K. (2016). Space weathering on airless bodies. Journal of Geophysical Research: Planets, 121(10), 1865-1884. https://doi.org/10.1002/2016JE005128

Pilorget, C., Fernando, J., Ehlmann, B. L., Schmidt, F., & Hiroi, T. (2016). Wavelength dependence of scattering properties in the VIS–NIR and links with grain-scale physical and compositional properties. Icarus, 267, 296–314. https://doi.org/10.1016/j.icarus.2015.12.029

Robinson, M. S., Brylow, S. M., Tschimmel, M., Humm, D., Lawrence, S. J., Thomas, P. C., Denevi, B. W., Bowman-Cisneros, E., Zerr, J., Ravine, M. A., Caplinger, M. A., Ghaemi, F. T., Schaffner, J. A., Malin, M. C., Mahanti, P., Bartels, A., Anderson, J., Tran, T. N., Eliason, E. M., … Hiesinger, H. (2010). Lunar reconnaissance orbiter camera (LROC) instrument overview. Space Science Reviews, 150(1–4), 81–124. https://doi.org/10.1007/S11214-010-9634-2

Rüsch, O., & Wöhler, C. (2022). Degradation of rocks on the moon: Insights on abrasion from topographic diffusion, LRO/NAC and Apollo images. Icarus, 115088. https://doi.org/10.1016/j.icarus.2022.115088

Rüsch, O., Marshal, R. M., Iqbal, W., Pasckert, J. H., van der Bogert, C. H., Patzek, M., (2022). Catastrophic rupture of lunar rocks: Implications for lunar rock size-frequency distributions, Icarus, 387. https://doi.org/10.1016/j.icarus.2022.115200

Rüsch, O., Sefton-Nash, E., Vago, J. L., Küppers, M., Pasckert, J. H., Khron, K., & Otto, K. (2020). In situ fragmentation of lunar blocks and implications for impacts and solar-induced thermal stresses. Icarus, 336. https://doi.org/10.1016/j.icarus.2019.113431

Sasaki, S., Nakamura, K., Hamabe, Y., Kurahashi, E., & Hiroi, T. (2001). Production of iron nanoparticles by laser irradiation in a simulation of lunar-like space weathering. Nature, 410(6828), 555-557. https://doi.org/10.1038/35069013

Sato, H., Robinson, M. S., Hapke, B., Denevi, B. W., & Boyd, A. K. (2014). Resolved Hapke parameter maps of the Moon. Journal of Geophysical Research E: Planets, 119(8), 1775–1805. https://doi.org/10.1002/2013JE004580

Schmidt, F., & Fernando, J. (2015). Realistic uncertainties on Hapke model parameters from photometric measurement. Icarus, 260, 73–93. https://doi.org/10.1016/j.icarus.2015.07.002

Shepard, M. K., & Campbell, B. A. (1998). Shadows on a Planetary Surface and Implications for Photometric Roughness. Icarus, 134(2), 279–291. https://doi.org/10.1006/icar.1998.5958

Shepard, M. K., & Helfenstein, P. (2007). A test of the Hapke photometric model. Journal of Geophysical Research E: Planets, 112(3). https://doi.org/10.1029/2005JE002625

Shkuratov, Y., Kaydash, V., & Videen, G. (2012). The lunar crater Giordano Bruno as seen with optical roughness imagery. Icarus, 218(1), 525–533. https://doi.org/10.1016/j.icarus.2011.12.023

Shkuratov, Y., Kaydash, V., Korokhin, V., Velikodsky, Y., Opanasenko, N., & Videen, G. (2011). Optical measurements of the Moon as a tool to study its surface. Planetary And Space Science, 59(13), 1326-1371. https://doi.org/10.1016/j.pss.2011.06.011

Shkuratov, Y., Stankevich, D. G., Petrov, D. V., Pinet, P. C., Cord, A. M., Daydou, Y. H., & Chevrel, S. D. (2005). Interpreting photometry of regolith-like surfaces with different topographies: Shadowing and multiple scattering. Icarus, 173(1), 3–15. https://doi.org/10.1016/j.icarus.2003.12.017

Shiltz, D. J., Bachmann, C. M., 2023. An alternative to Hapke's macroscopic roughness correction.

Icarus 390, 115240. https://doi.org/10.1016/j.icarus.2022.115240.

Shkuratov, Y. G., Helfenstein, P., 2001. The Opposition Effect and the Quasi-fractal Structure of Regolith. I. Theory. Icarus 152 (1), 96–116. https://doi.org/10.1006/icar.2001.6630.

Shkuratov, Y., Kaydash, V., Korokhin, V., Velikodsky, Y., Petrov, D., Zubko, E., Stankevich, D.,Videen, G., 2012. A critical assessment of the Hapke photometric model. Journal of Quantitative Spectroscopy and Radiative Transfer 113 (18), 2431–2456. https://doi.org/10.1016/j.jqsrt.2012.04.010.

Sides, S. C., Becker, T. L., Becker, K. J., Edmundson, K. L., Backer, J. W., Wilson, T. J., et al. (2017). The USGS integrated software for imagers and spectrometers (ISIS3) instrument support, new capabilities, and releases. In 48th Lunar and Planetary Science Conference

Speyerer, E. J., Povilaitis, R. Z., Robinson, M. S., Thomas, P. C., & Wagner, R. V. (2016). Quantifying crater production and regolith overturn on the Moon with temporal imaging. Nature, 538(7624), https://doi.org/215-218. 10.1038/nature19829

Spudis, P. D., Hawke, B. R., Lucey, P. G., Taylor, G. J., & Stockstill, K. R. (1996, March). Composition of the ejecta deposits of selected lunar basins from Clementine elemental maps. In 27th Lunar and Planetary Science Conference.

Sun, L. and Lucey, P.G. (2022). Near-infrared spectroscopy of boulders with dust or patina coatings on the Moon: A two-layer radiative transfer model. Icarus, 387, p. 115204. https://doi.org/10.1016/j.icarus.2022.115204.

Sun, L., & Lucey, P. G. (2021). Unmixing mineral abundance and Mg# with radiative transfer theory: Modeling and applications. Journal of Geophysical Research: Planets, 126, e2020JE006691. https://doi.org/10.1029/2020JE006691

Tanabe, N., Cho, Y., Tatsumi, E., Ebihara, T., Yumoto, K., Michikami, T., ... & Sugita, S. (2021). Development of image texture analysis technique for boulder distribution measurements: Applications to asteroids Ryugu and Itokawa. Planetary and Space Science, 204, 105249. https://doi.org/10.1016/j.pss.2021.105249

van Ginneken, B., Marigo Stavridi, and Jan J. Koenderink, "Diffuse and Specular Reflectance from Rough Surfaces," Appl. Opt. 37, 130-139 (1998)

Velichko, S., Korokhin, V., Shkuratov, Y., Kaydash, V., Surkov, Y., & Videen, G. (2022). Photometric analysis of the Luna spacecraft landing sites. Planetary and Space Science, 216, 105475. https://doi.org/10.1016/j.pss.2022.105475

Velichko, S., Korokhin, V., Velikodsky, Y., Kaydash, V., Shkuratov, Y., & Videen, G. (2020). Removal of topographic effects from LROC NAC images as applied to the inner flank of the crater Hertzsprung S. Planetary and Space Science, 193. https://doi.org/10.1016/j.pss.2020.105090

Velikodsky, Y. I., Opanasenko, N. V., Akimov, L. A., Korokhin, V. V., Shkuratov, Y., Kaydash, V., ... & Berdalieva, N. E. (2011). New Earth-based absolute photometry of the Moon. Icarus, 214(1), 30-45. https://doi.org/10.1016/j.icarus.2011.04.021

Verbiscer, A. J., & Veverka, J. (1989). Albedo dichotomy of Rhea: Hapke analysis of Voyager photometry. Icarus, 82(2), 336-353. https://doi.org/10.1016/0019-1035(89)90042-0

Walsh, J.M., Shreffler, R.G., Willig, F.J., 1953. Limiting conditions for jet formation in high velocity collisions. J. Appl. Phys. 24, 349–359. http://dx.doi.org/10.1063/1. 1721278.

Watkins, R. N., Jolliff, B. L., Boyd, A., Gonzales, N., & Speyerer, E. (2021). LROC NAC Photometry of the Apollo 16 Landing Site: Correlating Feldspathic Compositions using Landing Site and Sample Data. 52nd Lunar and Planetary Science Conference.

Watkins, R. N., Jolliff, B. L., Mistick, K., Fogerty, C., Lawrence, S. J., Singer, K. N., & Ghent, R. R. (2019). Boulder Distributions Around Young, Small Lunar Impact Craters and Implications for Regolith Production Rates and Landing Site Safety. Journal of Geophysical Research: Planets, 124(11), 2754–2771. https://doi.org/10.1029/2019JE005963

Wei, G., Byrne, S., Li, X., & Hu, G. (2020). Lunar surface and buried rock abundance retrieved from Chang'E-2 microwave and diviner data. The Planetary Science Journal, 1(3), 56. https://doi.org/10.3847/PSJ/abb2a8

Wu, Y., Gong, P., Liu, Q., & Chappell, A. (2009). Retrieving photometric properties of desert surfaces in China using the Hapke model and MISR data. Remote Sensing of Environment, 113(1), 213–223. https://doi.org/10.1016/J.RSE.2008.09.006

Zhou, P. et al. (2021). Retrieval of photometric parameters of minerals using a self-made multi-angle spectrometer based on the Hapke radiative transfer model. Remote Sensing, 13(15), p. 3022. https://doi.org/10.3390/rs13153022.

# Photometry of LROC NAC resolved rock-rich regions on the Moon
# Figures and Tables

Table 1: Table summarising the variables used in the default reflectance modelling approach.
The variables that are direct input to the Hapke reflectance equation (eq 2) are listed first and indicated with the square bracket.
The variables rock Size Frequency Distribution and aspect ratio of the rock are used to construct the artificial terrain models of the rock-rich regions. The cumulative fractional area of the LROC NAC resolved boulder fields is done by fitting a circle around individual boulders within a boulder field and calculating their resultant cumulative area.

| Parameter | Value | Rationale | Description |
|---|---|---|---|
| $b$ | 0.33 | Sato (2014) | Describes the angular width of the scattering lobe in the Henyey Greenstein function. |
| $c$ | $3.29 exp\ -(17.4b^2) - 0.908$ | Negative $c$ values: more backscattering; Positive $c$ values: more forward scattering (Hapke, 2012) | Describes the amplitude of the scattering lobe in the Henyey Greenstein function. |
| $\omega_{regolith}$ | 0.48 | Watkins et al. (2016) | The fraction of incident light scattered by the regolith. |
| $\omega_{rock}$ | 0.65 | Basaltic slab – De Angelis et al. (2017) | The fraction of incident light scattered by the rock surface. |
| Rock SFD | i) Exponential  ii) Double rock configuration | See Section 2.2 | Modeled distribution of size-frequency of rocks populating the surface surrounding young impact craters. |
| Aspect Ratio of Rock | 0.57 | Demidov and Basilevsky (2014) | Ratio of width to height of rocks. |
| Cumulative fractional Area of rocks (CFA) | Measured in LROC NAC images | - | Rock abundance within the sampled boulder field expressed as a summation of areas occupied by individual rocks. |

Table 2: Table of LROC NAC observations used in this study along with the viewing geometry details. Entries in white correspond to 'configuration 1' or "emission ratio" whereas the entries in gray correspond to 'configuration 2' or "incidence ratio ".The images used for the study at Hertzsprung S have been topographically corrected by Velichko et al. 2020. The images used for the study at North Ray are then resampled to a pixel resolution of ~1m in order to match the resolution of the DEM that was used to correct the images for regional topography.

| Image ID | Incidence Angle (degrees) | Emission Angle (degrees) | Phase Angle (degrees) | Resolution (m/pixel) | Location |
|---|---|---|---|---|---|
| M191466684L | 54.9 | 15.58 | 39.4 | 1.181 | Crater on the inner flank of Hertzsprung S |
| M191473833L | 53.9 | 1.75 | 55.72 | 1.145 | |
| M129187331R | 54.26 | 1.13 | 53.12 | 0.485 | North Ray crater, Apollo 16 landing site |
| M177535538R | 69.7 | 0.92 | 70.61 | 0.468 | |
| M129187331L | 54.19 | 1.69 | 55.86 | 0.486 | |
| M177535538L | 69.64 | 3.72 | 73.36 | 0.469 | |

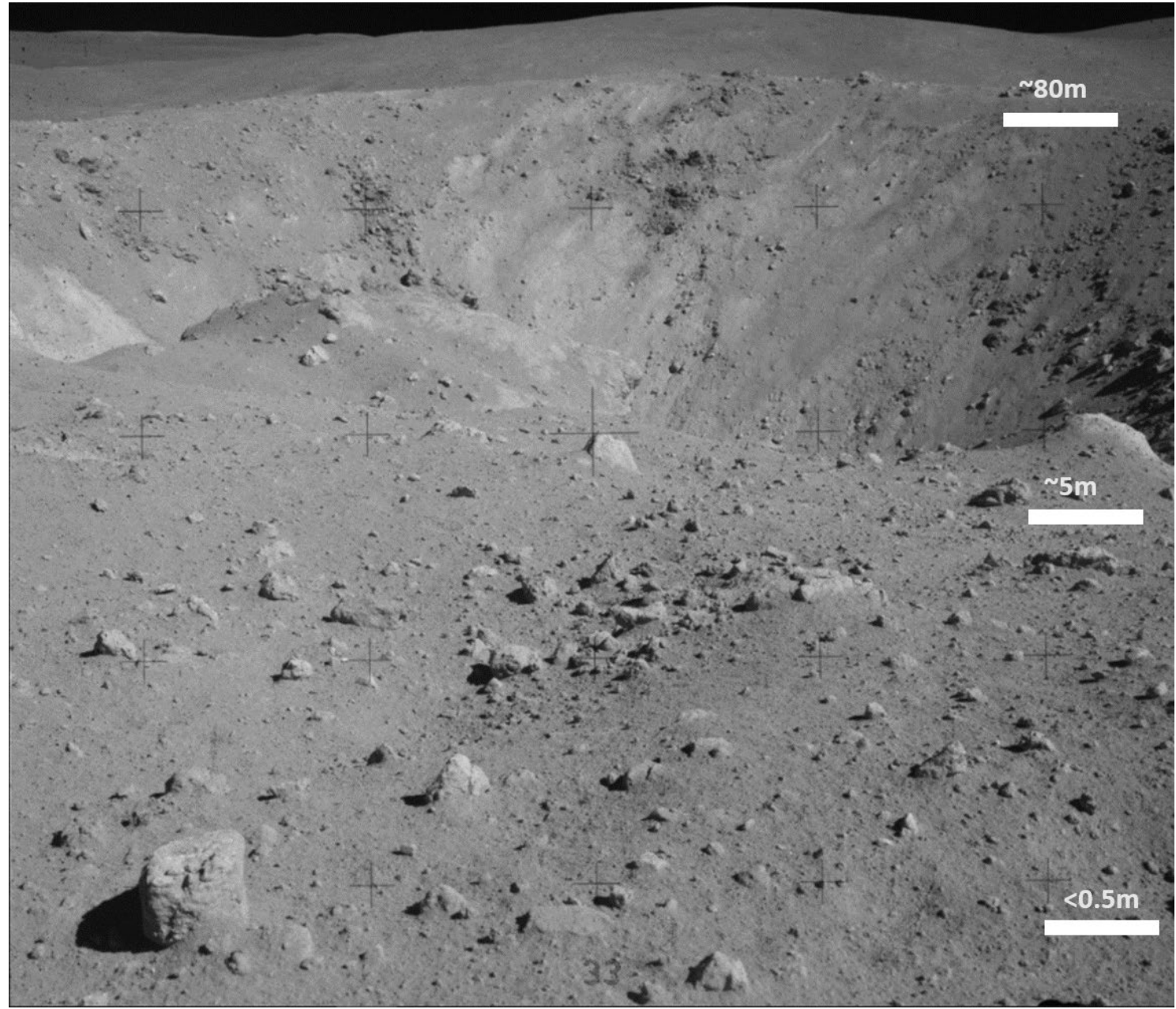

Figure 1 : A subset of an Apollo 16 Astronaut panoramic image original image ID: AS16-106-17256HR) depicting boulderfields on the rim of North Ray crater. Various rock morphologies exist within a local boulderfield. Simplistic morphologies such as the rock on the lower left in the foreground (similar to the non abraded profile in figure 2) exist while more varied morphologies are seen in the background.

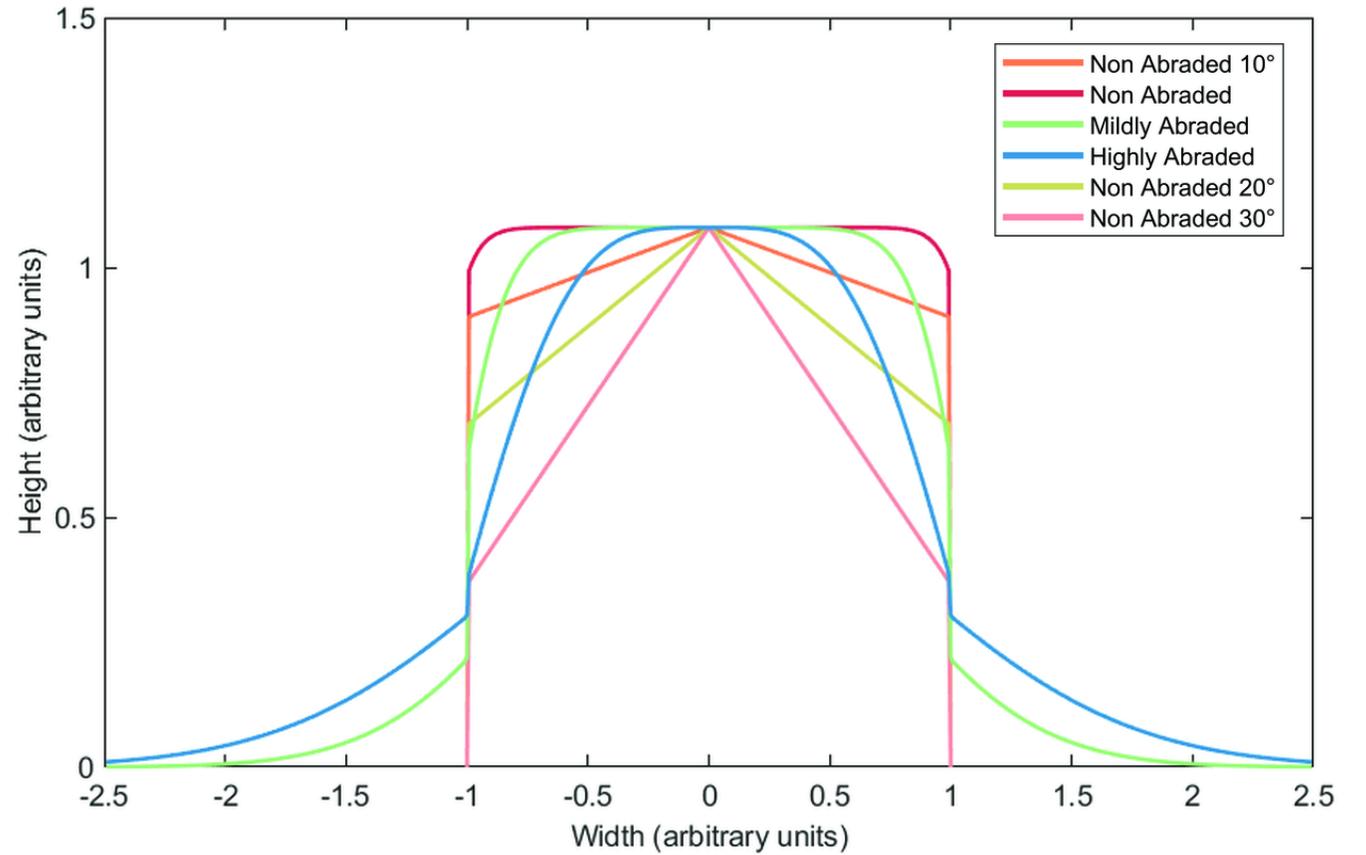

Figure 2 :Modelled DTM rock profiles. Units are arbitrary. These rock profiles are geologically informed and indicate various stages of abrasion (Rüsch and Wöhler, 2022) due to micrometeorite abrasion that caused rounding of edges and the development of a fillet. The rocks with tilted top faces present blocky or "angularity" is indicative of catastrophic disruption of the parent rock.

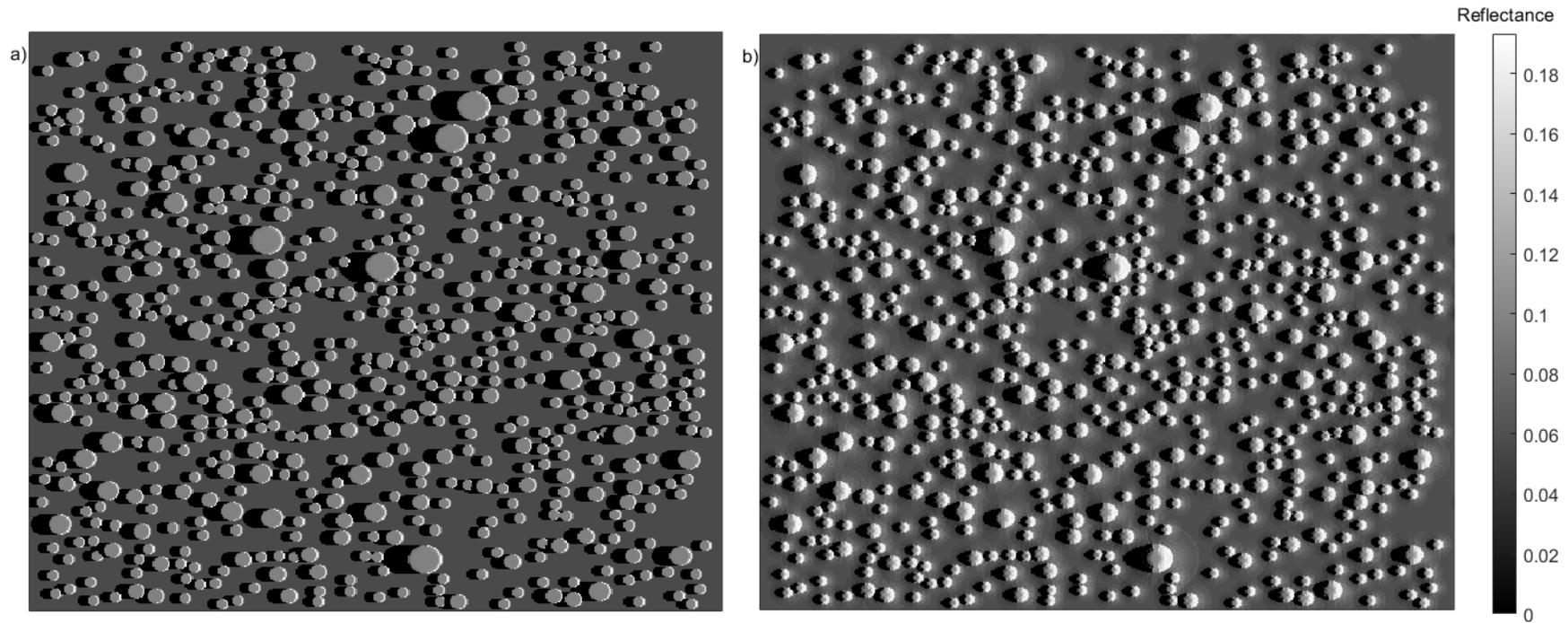

Figure 3: Artificial image for two rock morphologies namely: a) non-abraded and b) highly-abraded. The cumulative fractional area of rocks within these artificial surfaces is 25%. The images are modelled at 55° incidence angle and nadir emission. The rocks follow an exponential rock SFD (eq 1), and are modelled as backscattering with : $\omega_{rock}$=0.65 $\omega_{regolith}$=0.48 and input $\bar{\theta}$ for rock and regolith set to 0°.

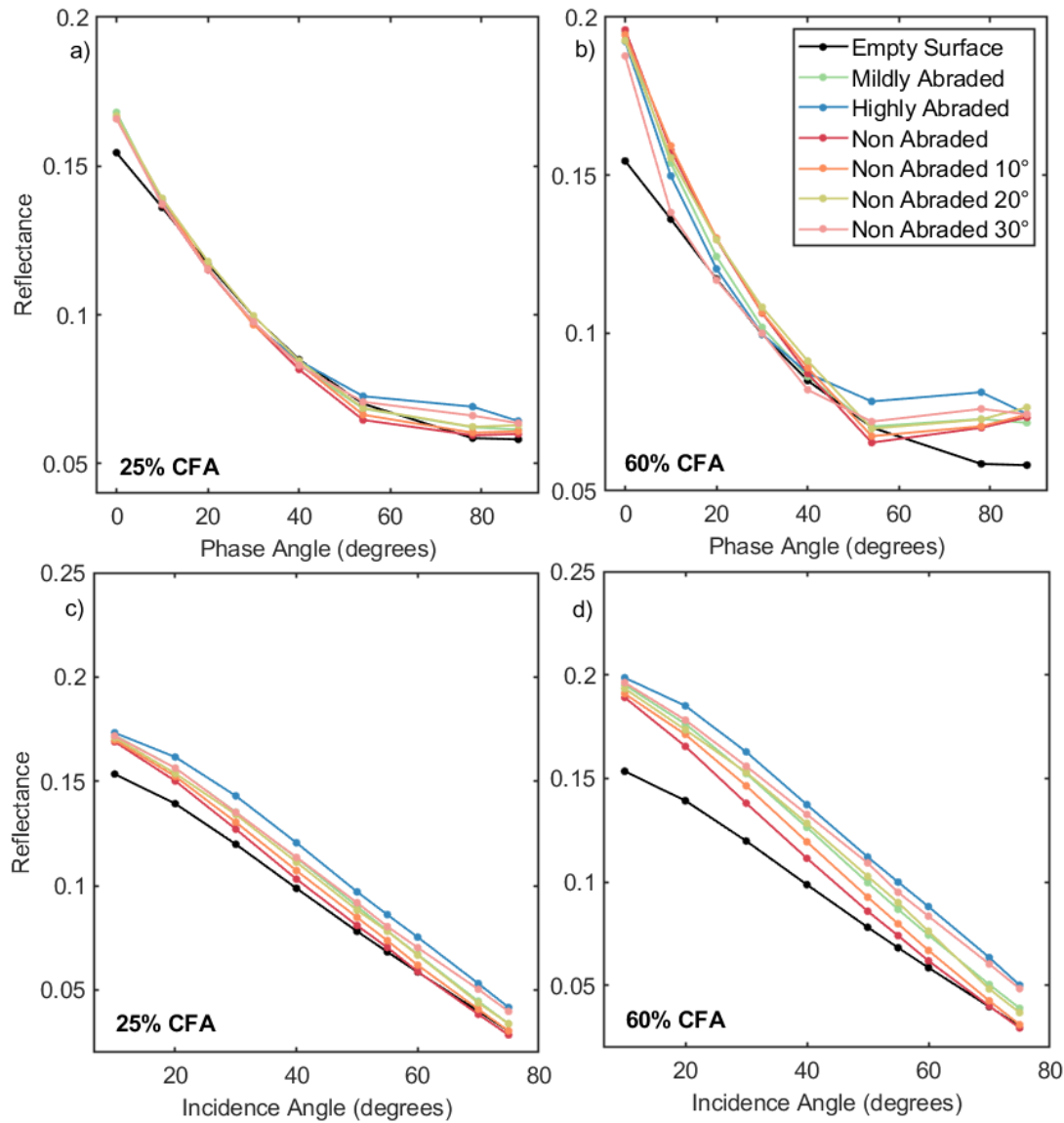

Figure 4: Variations in reflectance of simulated rock-rich surfaces across a) & b) phase angles and c) & d) incidence angles. Phase angle (configuration 1) versus mean modelled reflectance at a) 25% rock abundance and b) 60% rock abundance. Incidence angle (configuration 2) versus mean modelled reflectance at c) 25% rock abundance d) 60% rock abundance A surface that does not contain rocks i.e., the empty surface is also modelled and shown in black to compare the changes in the slope the reflectance vs phase angle/incidence angle curve and the absolute values of reflectance of rock rich surfaces to surfaces that do not host rocks.

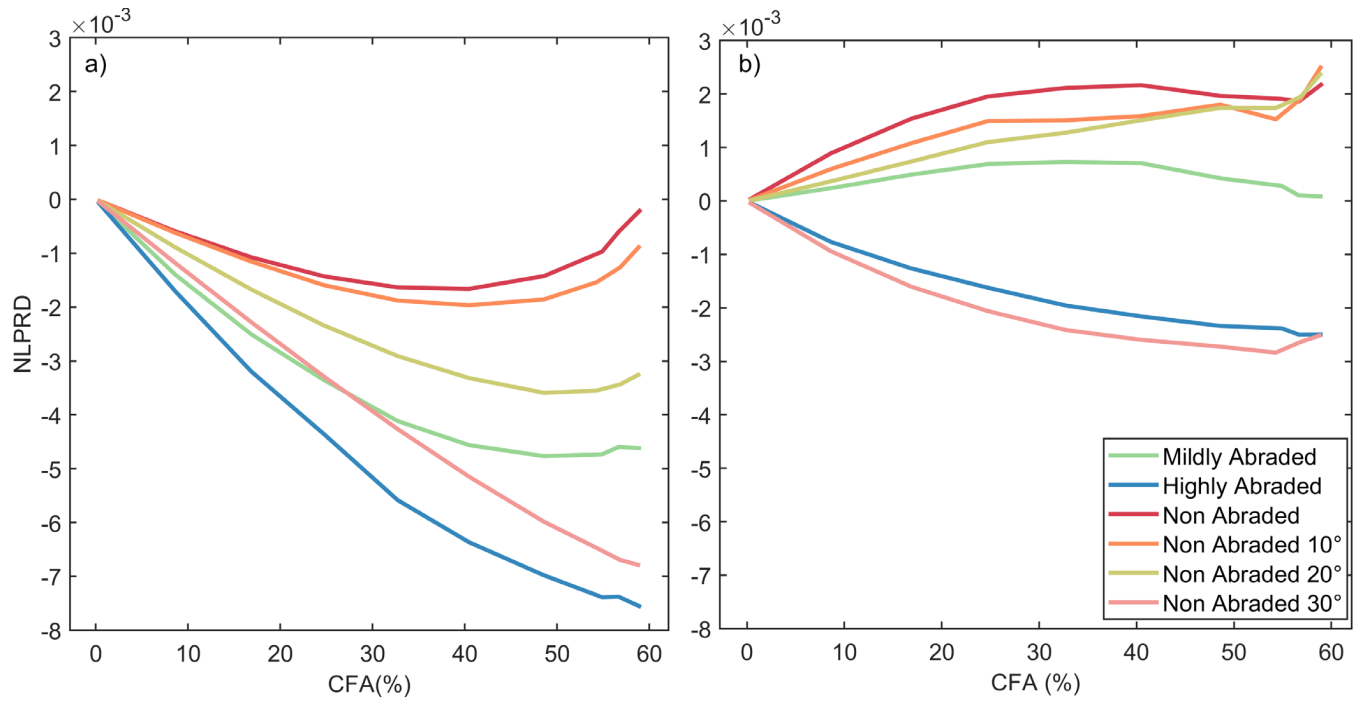

Figure 5: NLPRD as a function of rock abundance expressed as cumulative fractional area (CFA) for model topographies of various rock morphologies in the (a) "emission ratio - configuration 1" $i_1$=54°, $e_1$=2° and $i_2$= 55°, $e_2$= 16° and in the (b) "incidence angle - configuration 2" where $i_1$=55°, $i_2$= 70°

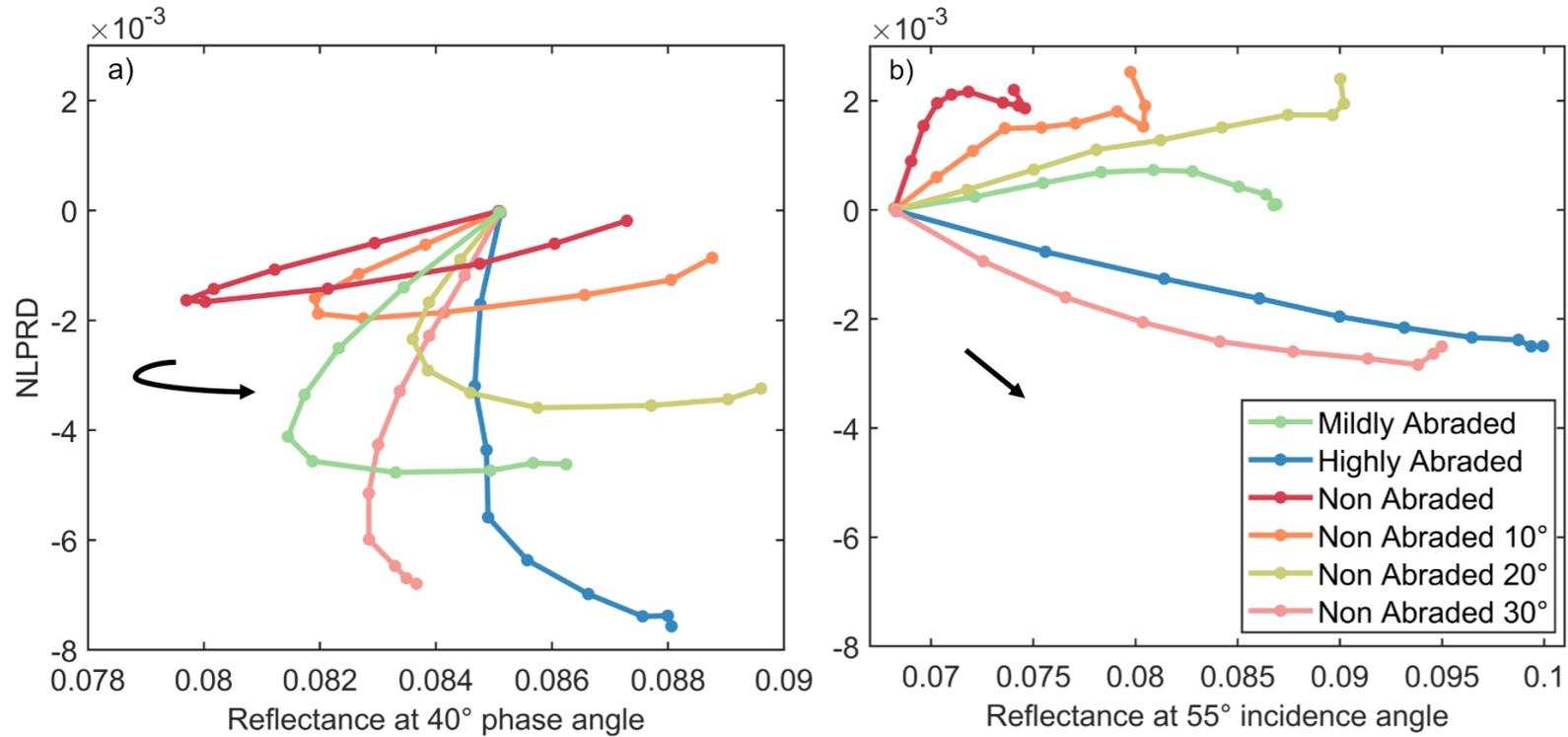

Figure 6: NLPRD as a function of reflectance for model topographies with increasing rock abundance in the (a) "emission ratio - configuration 1" and in the (b) "incidence angle - configuration .

a) NLPRD of modelled rock-rich regions are plotted against their mean reflectance at 40° phase angle b) NLPRD of modelled rock-rich regions are plotted against their mean reflectance at 55° phase angle. In both cases, the black arrows indicate the direction of increasing rock abundance (higher CFA)

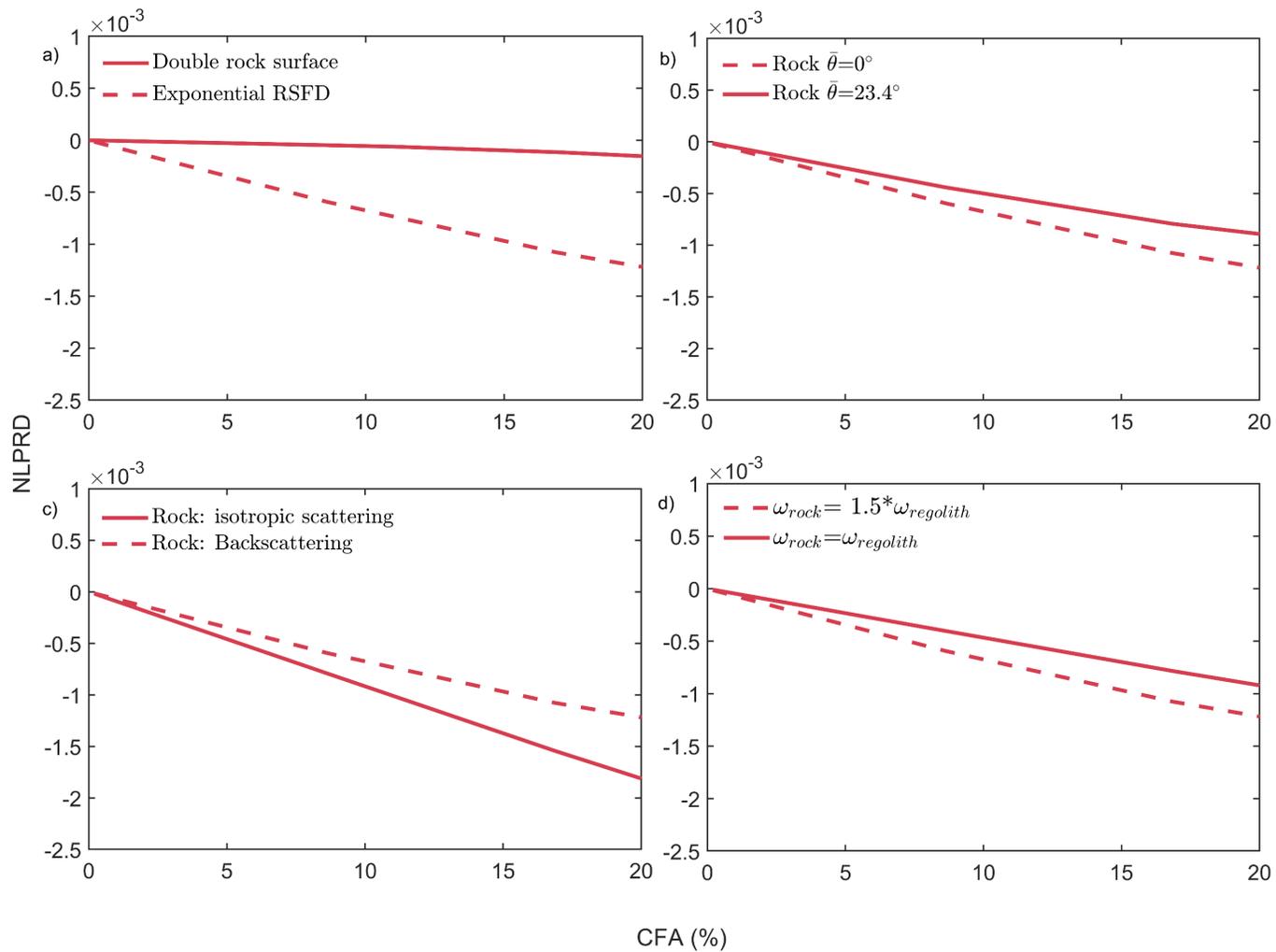

Figure 7: Effect of input variables on NLPRD values calculated on an emission ratio: a) Rock SFD, b) macroscopic roughness ($\bar{\theta}$), c) scattering parameters, and d) single scattering albedo ($\omega$) for rock. The default combination of parameters is denoted by the dashed red line and corresponds to: Morphology: Non-Abraded, scattering: Backscattering, rock SFD : Exponential (eq.1), $\omega_{rock}$=0.65, $\omega_{regolith}$=0.48, $\bar{\theta}$ =0°.

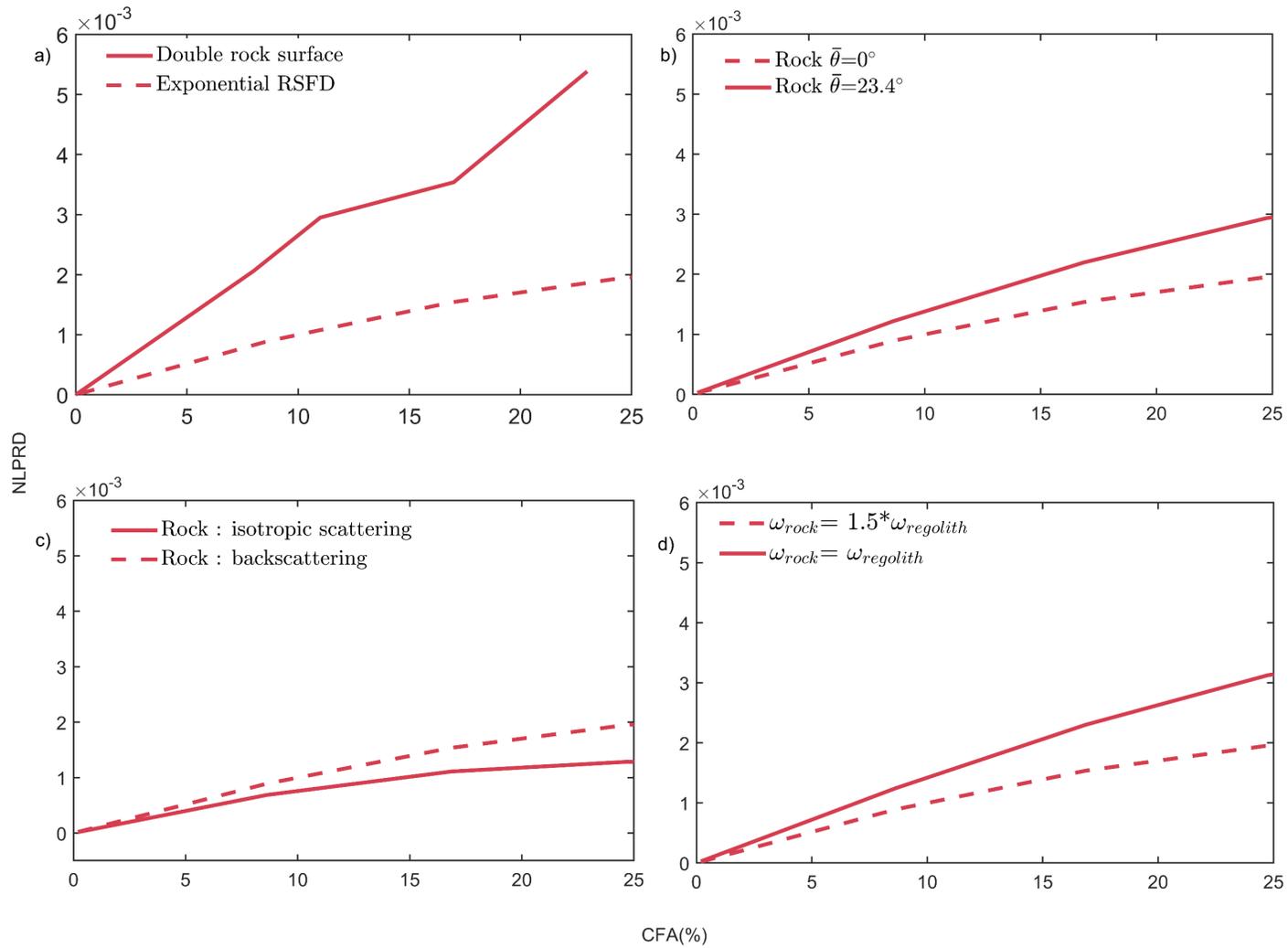

Figure 8: Effect of input variables on NLPRD values calculated on an incidence ratio: a) Rock SFD, b) macroscopic roughness ($\bar{\theta}$), c) scattering parameters, and d) single scattering albedo ($\omega$) for rock. The default combination of parameters is denoted by the dashed red line and corresponds to: Morphology: Non-Abraded, scattering: Backscattering, rockSFD : Exponential (eq.1), $\omega_{\text{rock}}$=0.65, $\omega_{\text{regolith}}$=0.48, $\bar{\theta}$ =0°.

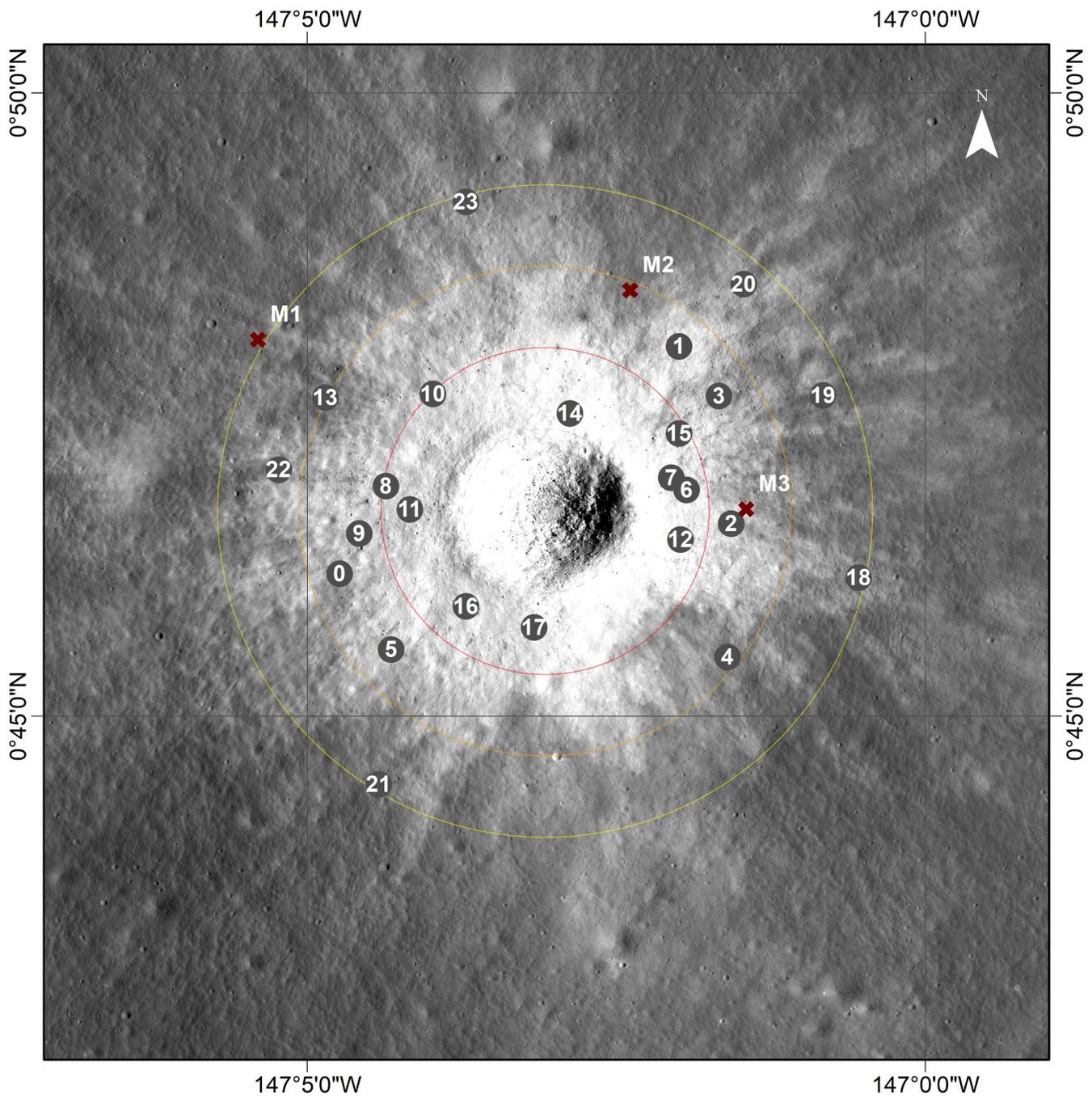

Figure 9: Overview of boulder fields polygons (numbered gray circles) and reference rock free (maroon crosses) at the unnamed crater on the inner flank of Hertzsprung S overlain on NAC image (M191466684L). The radial distance from the craters is denoted by the red, orange, and yellow circles that correspond to twice, thrice, and four times the crater radius respectively. The image resolution is ~1 m per pixel.

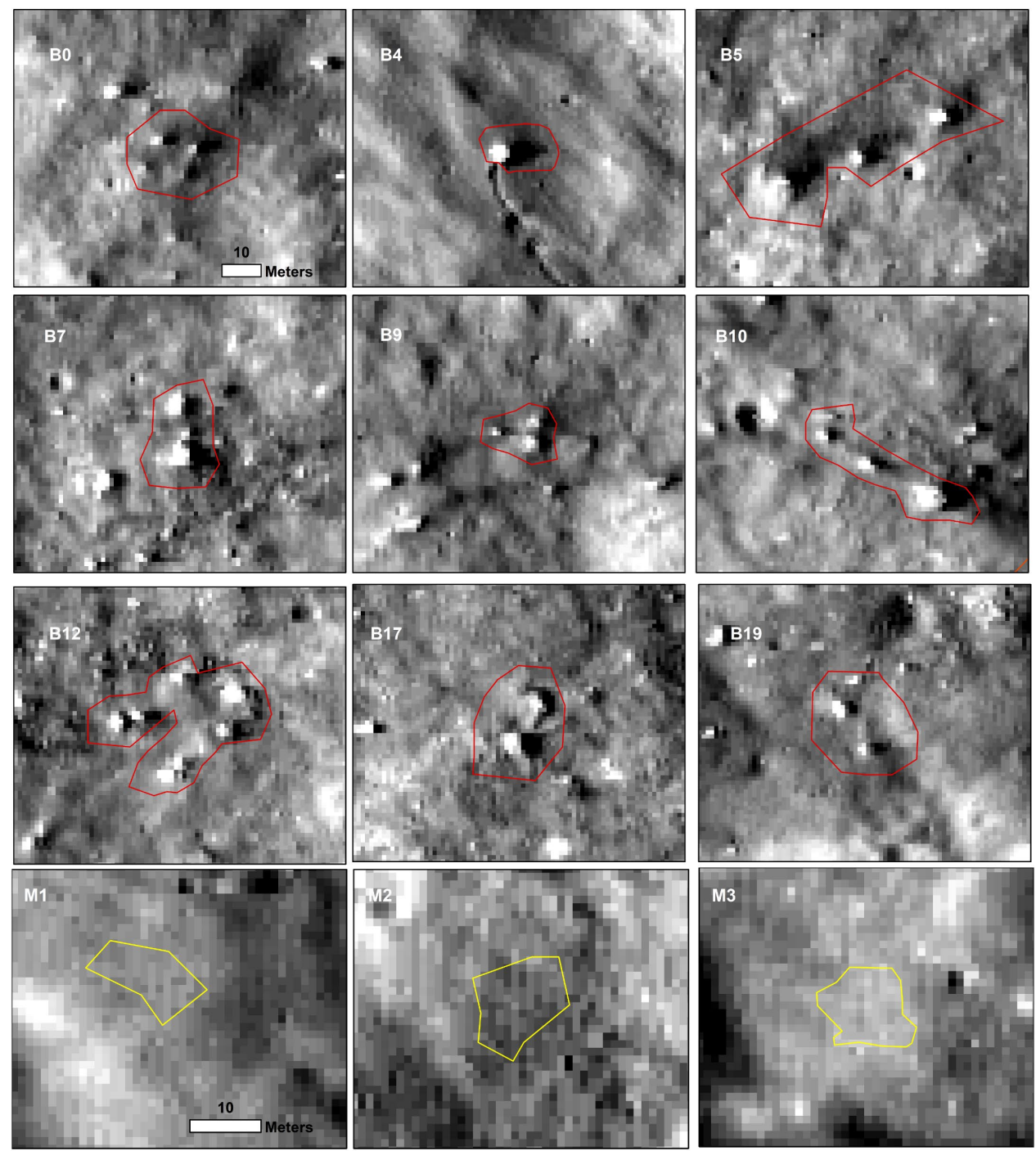

Figure 10: Zoom in view of a few of the boulder fields (in red) in the vicinity of the unnamed young crater on the inner flank of Hertzsprung S. The polygons in red indicate the extent to which the mean $A_{eq}$ is extracted post-topographic correction.

The last row shows the reference rock-free regions (in yellow) that are used to calculate the NLPRD of the boulder fields.

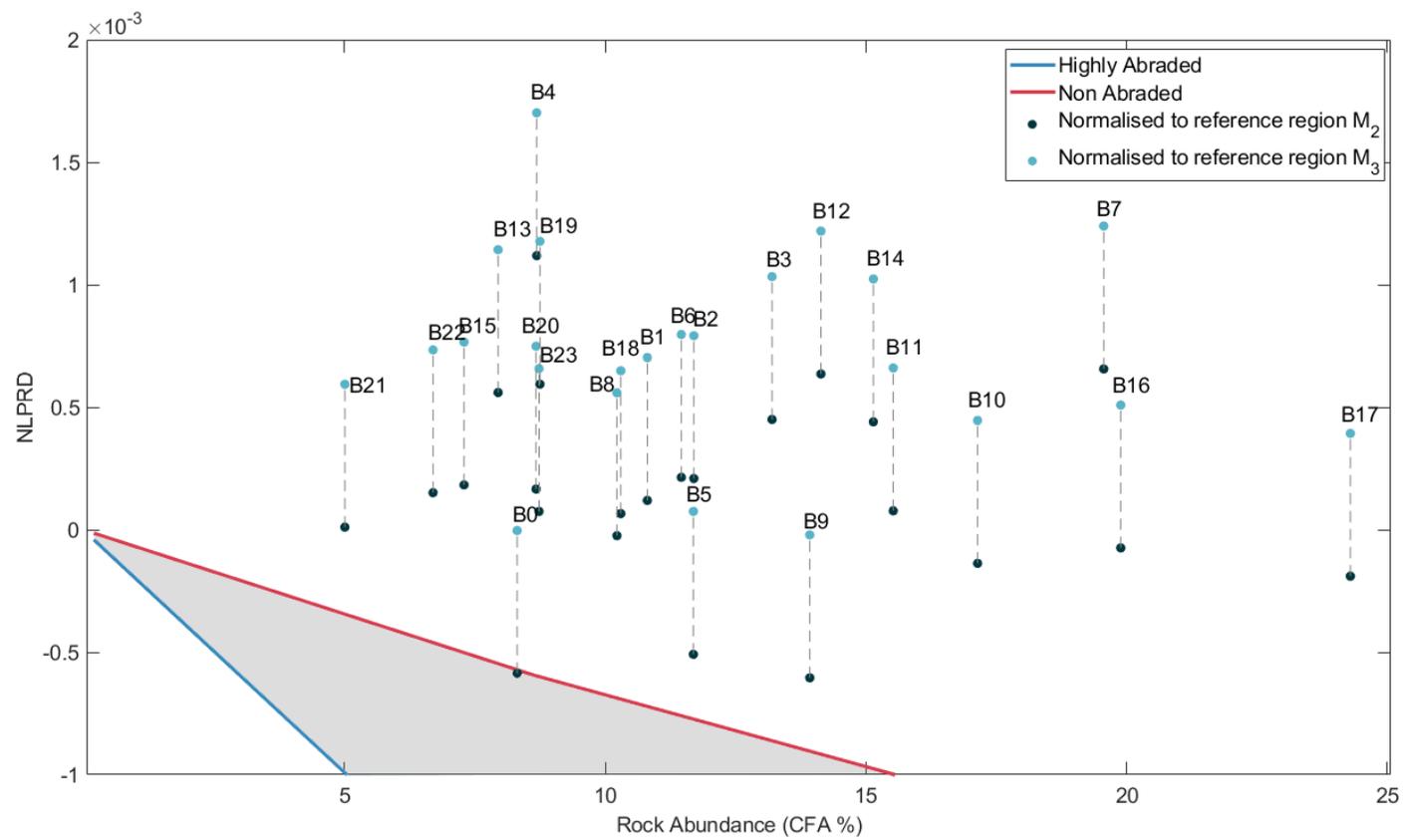

Figure 11: NLPRD of the boulder fields (shown in figure 10) in the vicinity of the unnamed crater at Hertzsprung S normalised to two reference rock-free surfaces (indicated by circles in blue and black) shown alongside the spread of model-derived NLPRD.

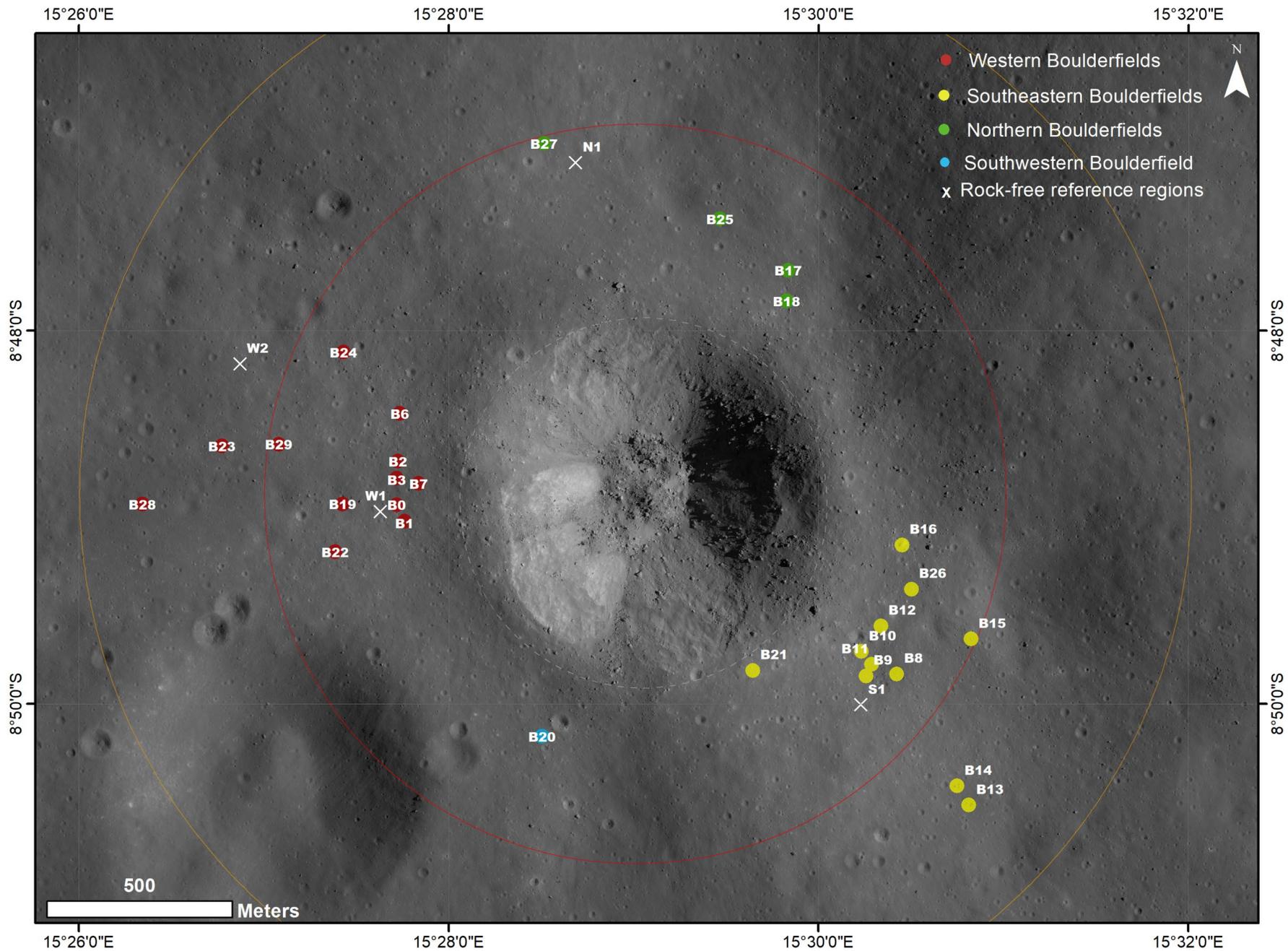

Figure 12: Overview of boulder fields indicated by the solid circles coloured by region and reference rock free (white crosses) at North Ray crater overlain on NAC images (M129187331L and M129187331R). The radial distance from the craters is denoted by the red and orange circles that correspond to twice and thrice the crater radius respectively. The image resolution is ~0.5 m per pixel.

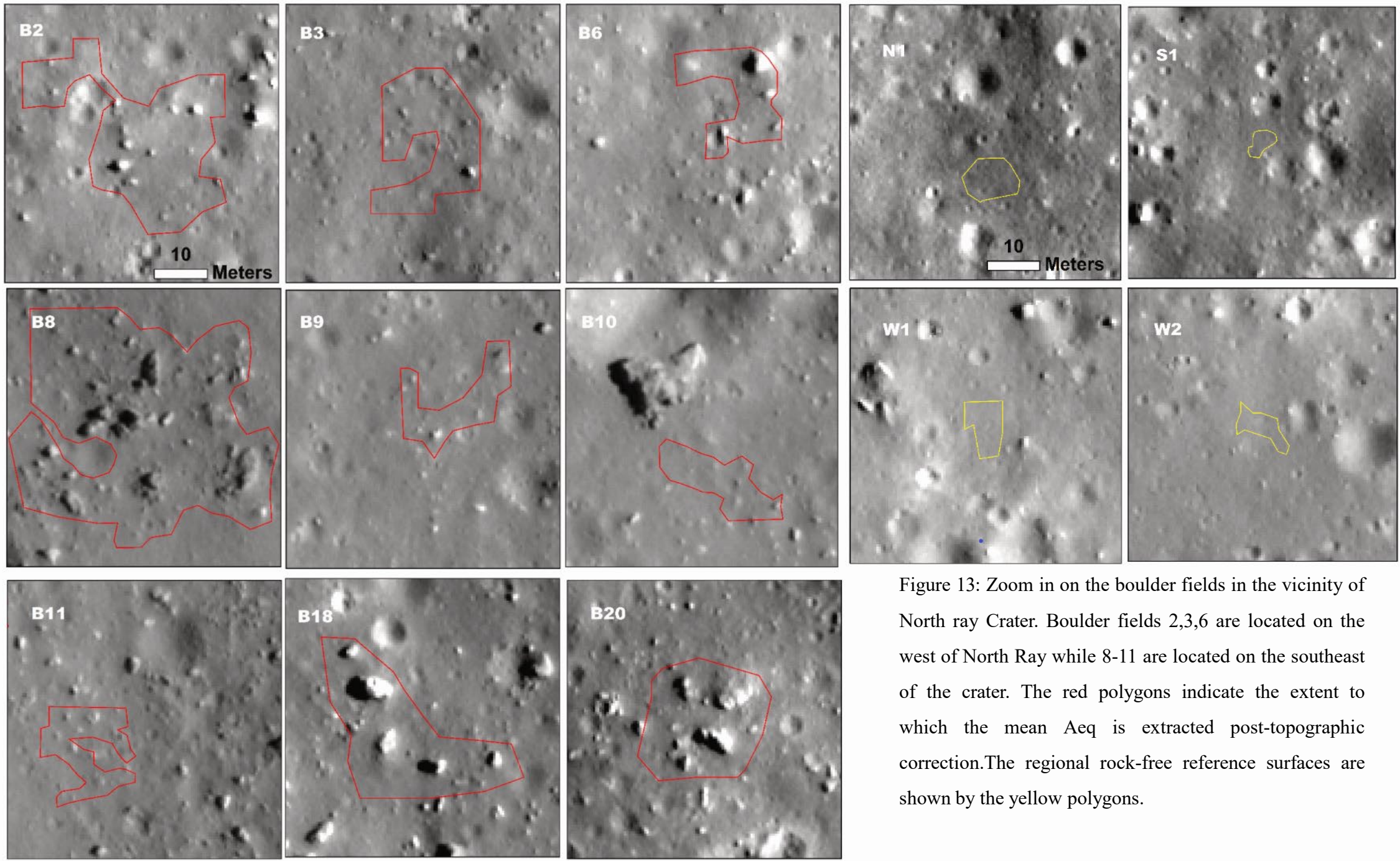

Figure 13: Zoom in on the boulder fields in the vicinity of North ray Crater. Boulder fields 2,3,6 are located on the west of North Ray while 8-11 are located on the southeast of the crater. The red polygons indicate the extent to which the mean Aeq is extracted post-topographic correction. The regional rock-free reference surfaces are shown by the yellow polygons.

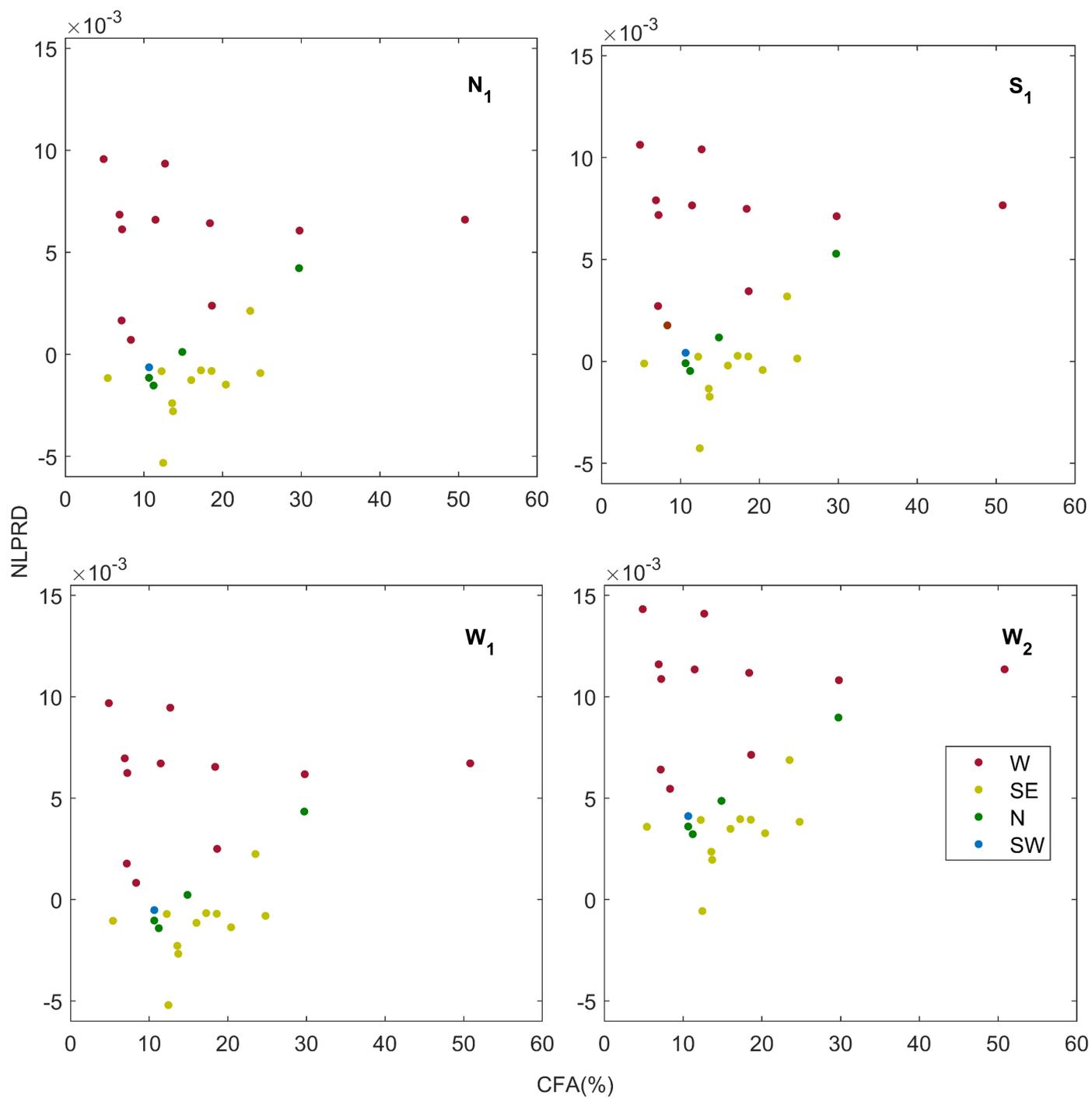

Figure 14: NLPRD of the boulder fields at North Ray with reference to four different rock-free reference regions (as shown in figure 13). The boulder fields are grouped according to their location with reference to the North Ray crater.

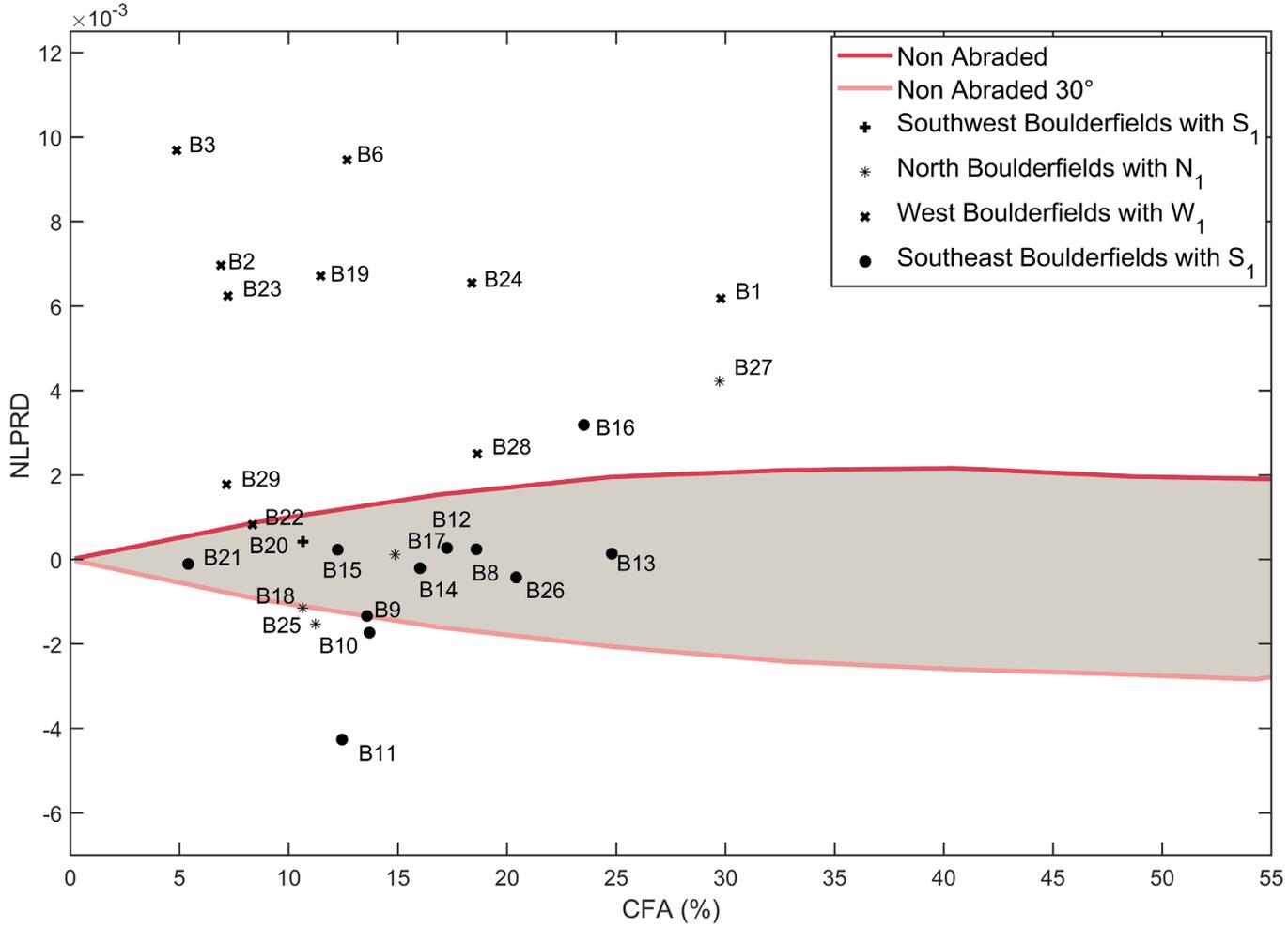

Figure 15: NLPRD of the boulder fields (shown in figure 13) at North Ray crater with normalised to their regional reference rock-free surfaces Additionally, the spread of the model derived NLPRD from fig 6a, is also plotted against rock abundance.